\documentclass[aps,prl,reprint,superscriptaddress]{revtex4-1}
\usepackage{graphicx,xr,natbib,hyperref,amsmath,amsthm,amssymb,wasysym,epstopdf,cleveref}
\usepackage{subfigure}
\usepackage[svgnames]{xcolor}
\hypersetup{hidelinks,colorlinks=true,allcolors=DarkBlue,pdfpagemode=UseNone}

\begin{document}
\newcommand{\figSize}{0.5}
\newcommand{\N}{^{14}\mathrm{N}}
\newcommand{\Hel}{H_\mathrm{el}}
\newcommand{\Ex}{|E_y\rangle}
\newcommand{\Ey}{|E_y\rangle}
\newcommand{\Eone}{|E_1\rangle}
\newcommand{\Etwo}{|E_2\rangle}
\newcommand{\Aone}{|A_1\rangle}
\newcommand{\Atwo}{|A_2\rangle}
\newcommand{\z}{|0\rangle}
\newcommand{\p}{|+1\rangle}
\newcommand{\m}{|-1\rangle}
\newcommand{\zN}{|0_N\rangle}
\newcommand{\pN}{|+1_N\rangle}
\newcommand{\mN}{|-1_N\rangle}

\title{Optical Control of a Single Nuclear Spin in the Solid State}

\author{M. L. Goldman}
\email[]{mgoldman@post.harvard.edu}
\affiliation{Department of Physics, Harvard University, Cambridge, Massachusetts 02138, USA}

\author{T. L. Patti}
\affiliation{Department of Physics, Harvard University, Cambridge, Massachusetts 02138, USA}

\author{D. Levonian}
\affiliation{Department of Physics, Harvard University, Cambridge, Massachusetts 02138, USA}

\author{S. F. Yelin}
\affiliation{Department of Physics, Harvard University, Cambridge, Massachusetts 02138, USA}
\affiliation{Department of Physics, University of Connecticut, Storrs, Connecticut 06269, USA}

\author{M. D. Lukin}
\affiliation{Department of Physics, Harvard University, Cambridge, Massachusetts 02138, USA}

\begin{abstract}

We demonstrate a novel method for coherent optical manipulation of individual nuclear spins in the solid state,  mediated by the electronic states of a proximal quantum emitter. Specifically, using
 the nitrogen-vacancy (NV) color center in diamond,  we demonstrate control of a proximal $^{14}$N nuclear spin via an all-optical Raman technique. We evaluate the extent to which the intrinsic physical properties of the NV center limit the performance of coherent control, and find that it is ultimately constrained by the relative rates of transverse hyperfine coupling and radiative decay in the NV center's excited state. Possible extensions and applications to other color centers are discussed.

\end{abstract}
\maketitle

Individual nuclear spins in solid-state materials feature exceptional isolation from the external environment, making them  promising candidates for applications such as long-lived quantum memory.  For example, the individual electronic spins associated with a nitrogen-vacancy (NV) center in diamond can couple coherently to both the nitrogen nucleus, which is most commonly the spin-1 $\N$ isotope, and to individual spin-1/2 $^{13}$C nuclei in the diamond lattice.   Even at room temperature, nuclear spins coupled to NV centers can be read out in a single shot \cite{Neumann2010} and can have coherence times longer than one second \cite{Maurer2012a}.  These properties have encouraged demonstrations of quantum registers consisting of the NV center's electronic spin and either the nitrogen nuclear spin \cite{Fuchs2011}, a single $^{13}$C nuclear spin \cite{Dutt2007}, or small networks of $^{13}$C nuclear spins \cite{Taminiau2014,Reiserer2016}.  Recently, these techniques have enhanced  NV-based magnetometry \cite{Lovchinsky2016} and enabled active quantum error correction \cite{Waldherr2014,Cramer2016,Hirose2016}. However, the substantial isolation of nuclear spins also complicates the initialization and coherent manipulation of nuclear spin states.  In particular, the above applications have generally relied on microwave and radio frequency fields for precise control of the electronic and nuclear spins, typically rendering them slow and of poor spatial resolution.

In this Letter, we demonstrate initialization, coherent manipulation, and readout of a single $\N$ nuclear spin using purely optical methods.  There has been extensive interest in using coherent optical techniques to manipulate the NV center's electronic spin, and to implement this spin control in increasingly sophisticated quantum operations \cite{Buckley2010,Yale2013,Chu2015,Yale2016,Sekiguchi2017,Zhou2017}. In some cases, these techniques are sensitive to the $\N$ nuclear spin \cite{Togan2011a,Golter2013,Golter2014a}.  All-optical approaches to spin manipulation are both typically faster and feature a much higher spatial resolution than the traditional spin manipulation techniques which use microwave or radio-frequency radiation. In particular, this makes optical manipulation ideal for densely spaced arrays of color centers integrated into nanophotonic devices, as has been demonstrated recently \cite{Evans2018}. We find that, rather than being limited by the small gyromagnetic ratio of the nuclear spin, the coherence and speed of this optical technique are limited by the coupling strength between the nuclear spin and the NV center's electronic spin.

We use optical Raman transitions wherein two ground states are optically coupled to a common excited state.  Specifically, we couple the states, labeled $\p$ and $\z$, of the NV center's electronic spin-triplet ground state manifold that have electronic spin projections $m_S=+1$ and $0$.  If the two optical transitions are driven far from resonance, then the states are coupled with an effective Rabi frequency $\tilde{\Omega} = \Omega_0^\ast \, \Omega_{+1}/\Delta,$ where $\Omega_{0(+1)}$ is the Rabi frequency of the transition from $\p$ ($\z$) and $\Delta$ is the transitions' single-photon detuning.  We must also consider the nuclear degree of freedom, such that our two ground states $|0,m_I^{(0)}\rangle$ and $|+1,m_I^{(+1)}\rangle$ are product states of specific electronic and nuclear spins.  Since the ground state of the NV center is an orbital singlet, the selection rules that determine the specific combinations of nuclear spin projections $m_I^{(0)}$ and $m_I^{(+1)}$ that can be coupled by this Raman technique are determined solely by the electronic and nuclear spin characteristics of the intermediate excited state.  In order to couple $|0,m_I^{(0)}\rangle$ to $|+1,m_I^{(+1)}\rangle$, the intermediate state must contain a superposition of the product states $|m_S=0,m_I=m_I^{(0)}\rangle$ and $|m_S=+1,m_I=m_I^{(+1)}\rangle$ so that the optical transitions, which cannot themselves flip either the electronic or nuclear spin, to both ground states are allowed.

Such a superposition of electronic and nuclear spin states is available as a result of an excited state level anticrossing (ESLAC), which has previously been explored in the contexts of nonresonant optical polarization of nearby nuclear spins \cite{Jacques2009,Steiner2010,Smeltzer2009} and early demonstrations of coherent population trapping of the electronic spin state \cite{Yale2013,Santori2006}.  As described in the Supplemental Materials \cite{SuppInfo}, two excited states with primarily $m_S=0$ and $m_S=+1$ character, which we respectively identify as $\Ey$ and $\Eone$, can be brought near degeneracy by the application of a magnetic field along the N-V axis.  This degeneracy is lifted by three interactions that couple $\Ey$ with $\Eone$ \cite{Doherty2011}.  The first two interactions, electronic spin-spin coupling and Zeeman coupling due to magnetic fields perpendicular to the N-V axis, mix $\Ey$ and $\Eone$ irrespective of the nuclear degree of freedom and enable Raman transitions that conserve nuclear spin ($m_I^{(0)}=m_I^{(+1)}$).  The third interaction, transverse hyperfine coupling with the $\N$ nuclear spin (which is much stronger in the excited state than in the ground state \cite{Fuchs2008,Steiner2010}), provides an electron-nuclear flip-flop interaction ($\propto S_+ I_- +\,  S_- I_+$)  and enables Raman transitions that conserve the total spin ($m_I^{(0)}=m_I^{(+1)}-1$).  Therefore, by driving optical transitions to $\Eone$ or $\Ey$ and setting the two-photon detuning $\delta_L$ between the applied optical fields equal to the frequency difference between two ground states, we can drive specific Raman transitions that conserve either the nuclear spin or the total spin.

\begin{figure}
\begin{center}
\subfigure{
\includegraphics[width=\columnwidth]{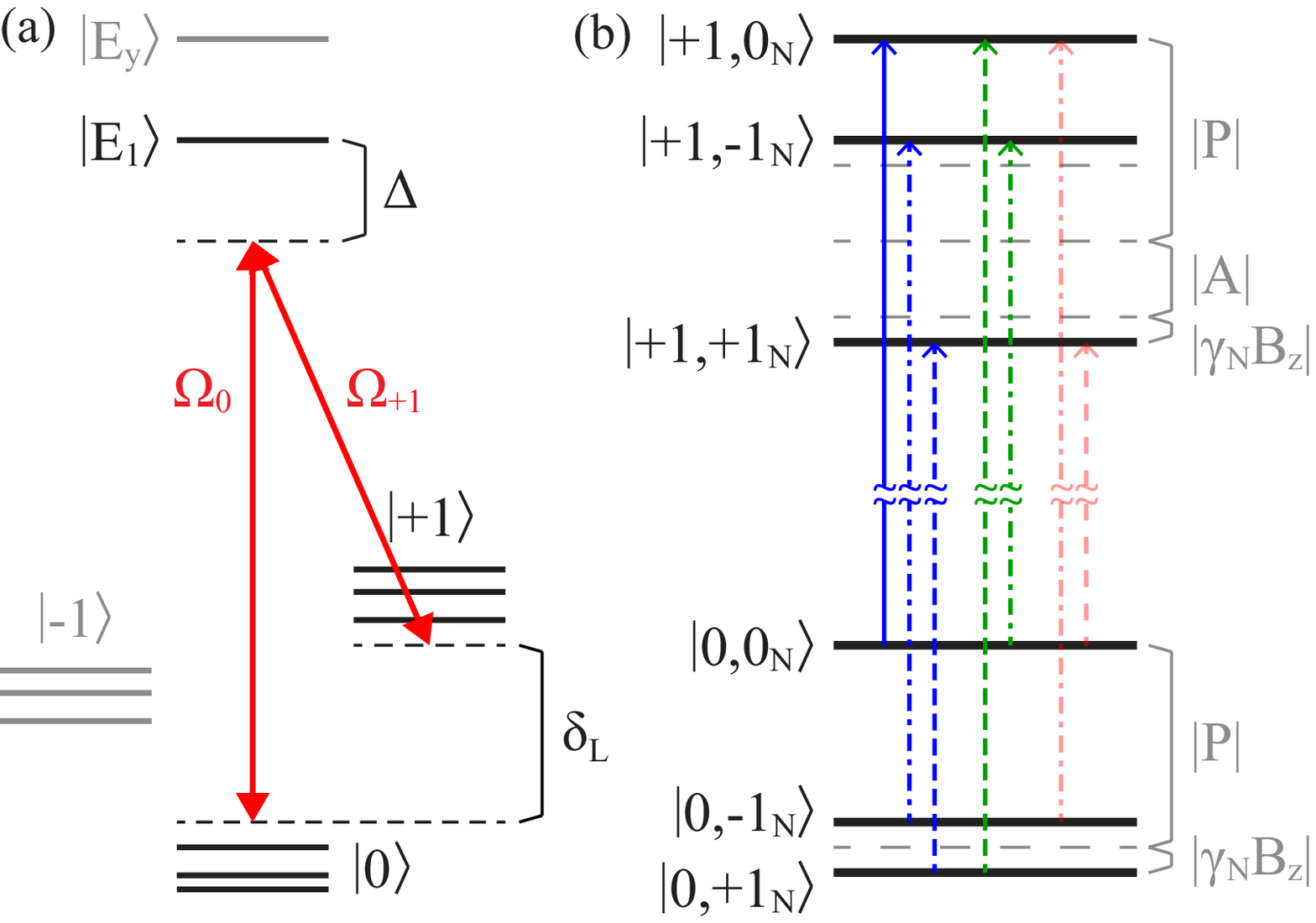}
}
\subfigure{
\includegraphics[width=\columnwidth]{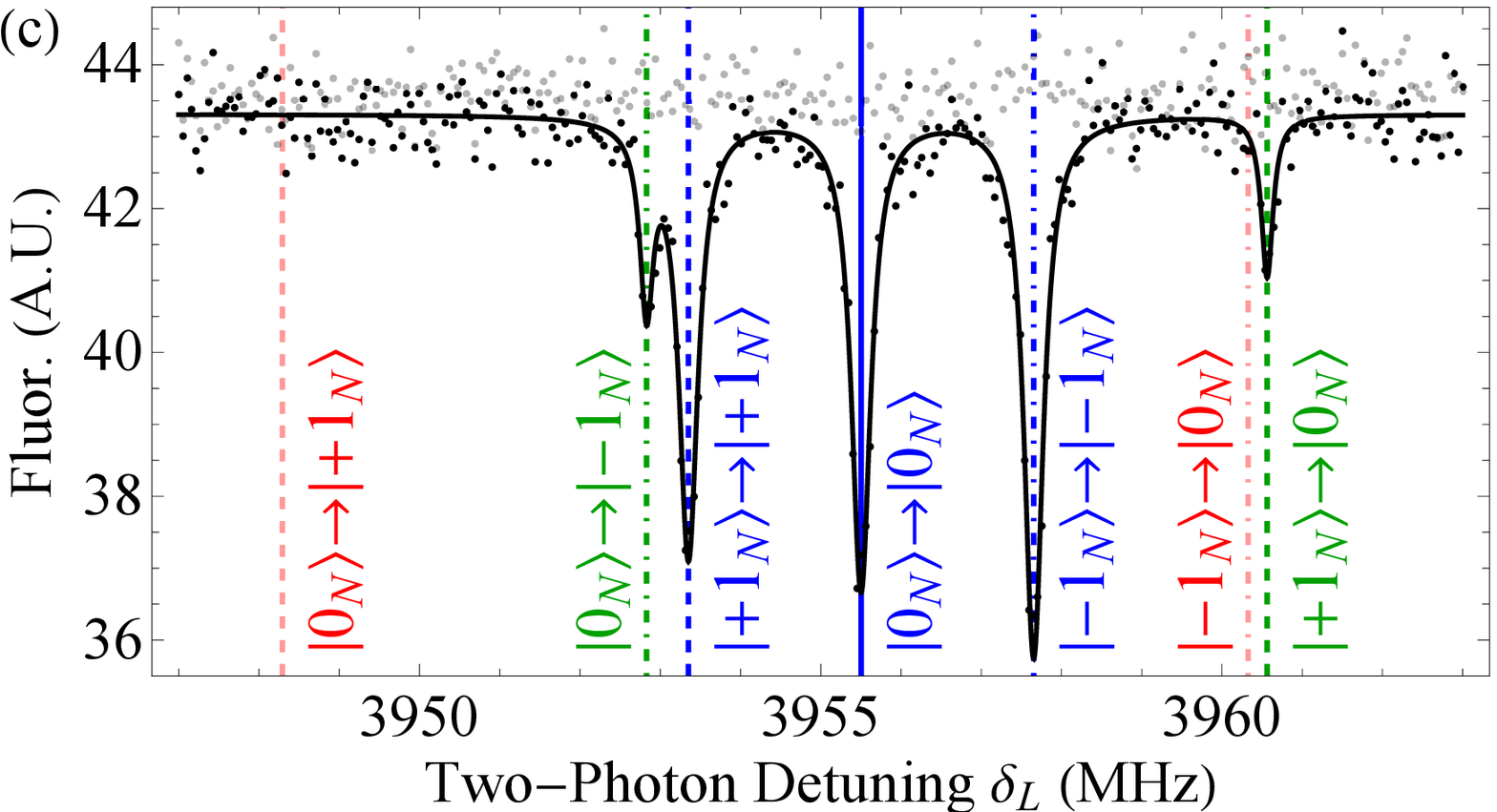}
}
\caption{(a) The Raman transition between the ground states $\z$ and $\p$, which is driven  via the optical excited states $\Eone$ and $\Ey$.  We indicate the two-photon detuning $\delta_L$, the Raman detuning $\Delta$, and the ground state hyperfine structure.  (b)  The hyperfine structure of the $\z$ and $\p$ electronic states, which depends on the $\N$ hyperfine coupling $A$, quadrupolar shift $P$, and gyromagnetic ratio $\gamma_N$.  We show the $\z\rightarrow\p$ transitions with $|\Delta m_I|\le 1$ that conserve the nuclear spin (blue), conserve the total spin (green), and do not conserve spin (red).  (c) Spectroscopy of the hyperfine structure of the $\z\rightarrow\p$ Raman transition, with vertical lines marking the fitted energies of the transitions shown in (b).  The light grey plot corresponds to a control experiment during which no Raman pulse was applied.  The subscript $N$, here and throughout the Letter, indicates the $\N$ spin.}
\label{fig.Raman ESR}
\end{center}
\end{figure}

To explore the nuclear-specific Raman transitions experimentally, we address a single NV center in the sample described in Ref. \cite{Goldman2015} (see also  Supplemental Materials \cite{SuppInfo}).  By holding the sample at a temperature of approximately 7 K, we observe 16 of the 18 possible optical transitions between $^3A_2$ and $^3E$ \cite{SuppInfo}, which we individually address using three external-cavity diode lasers at 637 nm that are gated by a combination of acousto-optic and electro-optic modulators.  The two transitions that are not observed, from $\z$ to $\Aone$ and $\Atwo$, are expected to be weaker because of those states' negligible $m_S=0$ admixture.  We drive a Raman transition from $\z$ to $\p$, as shown in \crefformat{figure}{Fig.~#2#1{(a)}#3}\cref{fig.Raman ESR}, by applying two optical driving fields that are detuned by $\Delta\approx870$ MHz below the transitions from $\z$ and $\p$ to $\Eone$.  Because these two fields are created by modulating a single laser with an electro-optic modulator, their relative phase is stable and their frequency difference $\delta_L$ can be controlled precisely.  By sweeping $\delta_L$, we map out the multiple hyperfine transitions between $\z$ and $\p$, which are depicted in \crefformat{figure}{Fig.~#2#1{(b)}#3}\cref{fig.Raman ESR}.  We perform this spectroscopy by initializing into the electronic $\z$ state, applying a Raman pulse to drive a transition from $\z$ to $\p$, and then reading out the population remaining in $\z$.

The results, shown in \crefformat{figure}{Fig.~#2#1{(c)}#3}\cref{fig.Raman ESR}, indicate that this Raman technique is sensitive to the state of the $\N$ nuclear spin.  We observe the three transitions, marked by blue lines, that represent flipping the electronic spin while preserving the nuclear spin, which are commonly observed using traditional microwave manipulation techniques \cite{SuppInfo}.  Crucially, we also observe two other transitions, marked by green lines, that represent driving the electron-nuclear flip-flop transition.  We do not observe other transitions, marked by faint red lines, that would conserve neither the total nor the nuclear angular momentum.  The relative frequencies of the five observed transitions are determined solely by the $\N$ axial hyperfine coupling strength $A$, quadrupole shift $P$, and axial Zeeman shift $\gamma_N B_z$ \cite{Doherty2012a}.  We extract the strength of the nuclear Zeeman shift $\gamma_N B_z/h = -118$ kHz from a measurement of the much larger electronic Zeeman splitting between $\p$ and $\m$ \cite{SuppInfo}, and the fitted values of $A/h = -2.151(4)$ MHz and $P/h = -4.942(9)$ MHz are in excellent agreement (less than 1\% deviation) with previous measurements \cite{Steiner2010,Smeltzer2009}.

\begin{figure}
\begin{center}
\subfigure{
\includegraphics[width=\columnwidth]{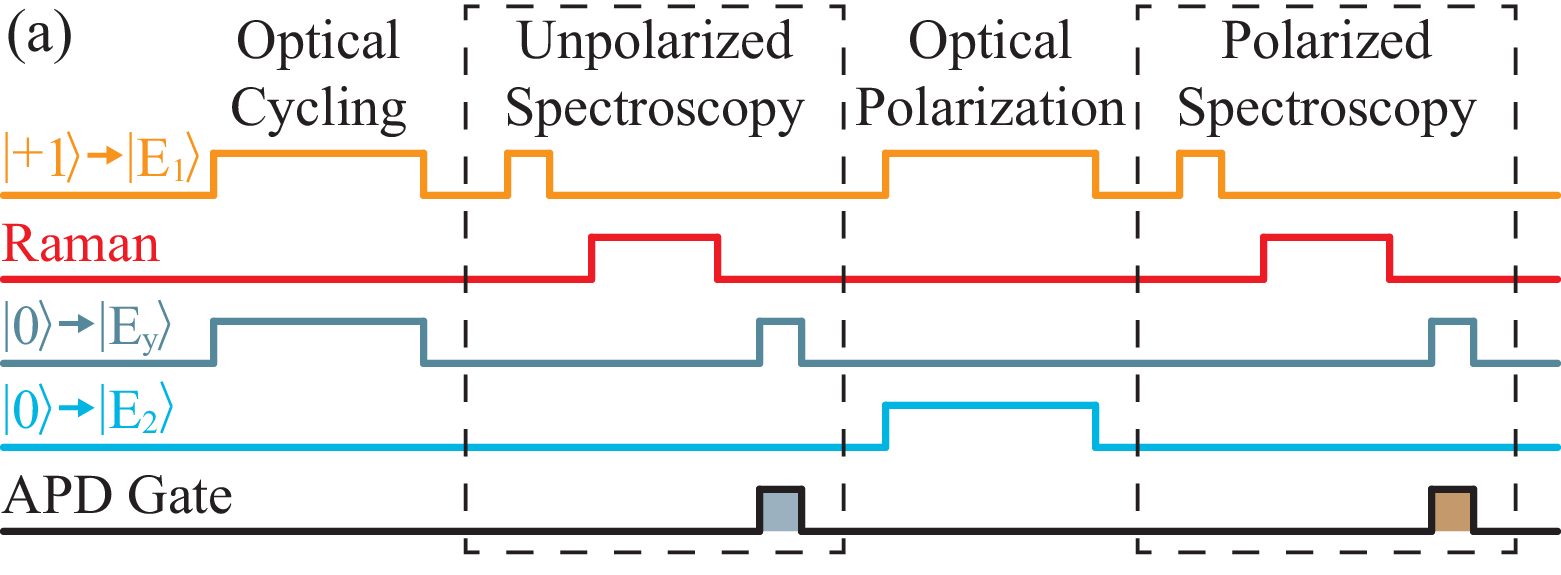}
}
\subfigure{
\includegraphics[width=\columnwidth]{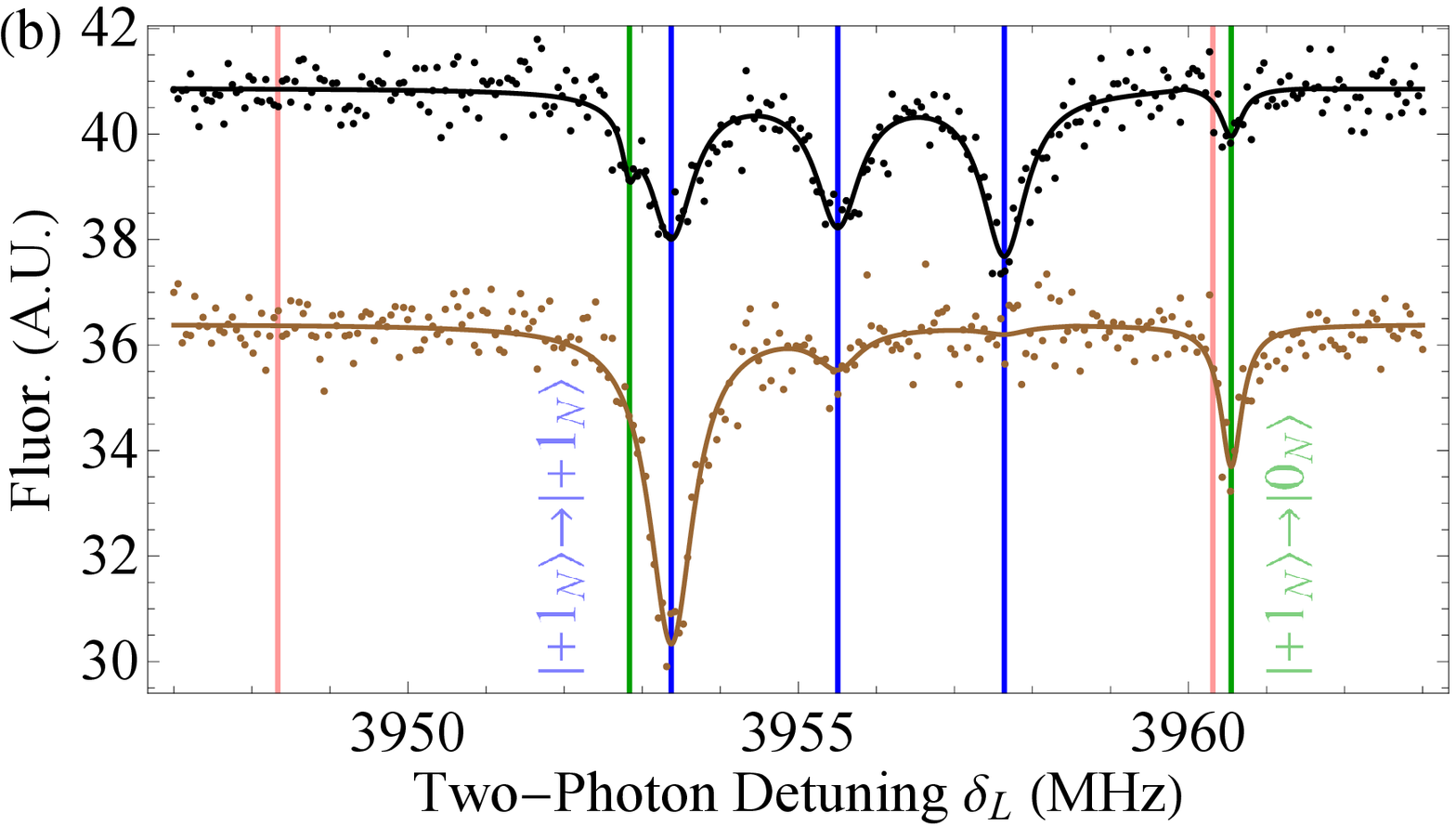}
}
\caption{Optical polarization of the $\N$ nuclear spin.  (a)  A schematic diagram of the pulse sequence used to measure the degree of nuclear polarization, which is described in more detail in the Supplemental Materials \cite{SuppInfo}.  (b) Spectroscopy of the $\z\rightarrow\p$ Raman transition, conducted before (black, offset vertically for clarity) and after (brown) 200 $\mu$s of optical polarization.}
\label{fig.Nuclear Polarization}
\end{center}
\end{figure}

Next, we describe a method for polarizing the $\N$ nuclear spin via optical pumping.  This method relies on simultaneously pumping on the $\p\rightarrow\Eone$ transition and the weakly allowed $\z\rightarrow\Etwo$ transition, both of which can induce a nuclear spin flip with $\Delta m_I = +1$ when the NV center decays to a different ground state \cite{SuppInfo}.  This method is distinct from the previously observed, nonresonant optical nuclear polarization mechanism \cite{Jacques2009,Steiner2010,Smeltzer2009} in that we can choose, by pumping on the appropriate transitions, to either polarize the nuclear spin or not while optically cycling the electronic spin.  We demonstrate this ability in Fig. \ref{fig.Nuclear Polarization}.  Using spectroscopy of the hyperfine Raman transitions, we show that the nuclear spin is largely unpolarized after an initial period of optical cycling, which we use to confirm that the NV center is in the correct charge state, and that we can subsequently polarize the nuclear spin by pumping on a different set of optical transitions.  The enhancement of the hyperfine transitions originating in $\pN$ and the suppression of all others implies a nuclear polarization of approximately $87\,\%$.

\begin{figure}
\begin{center}
\includegraphics[width=\columnwidth]{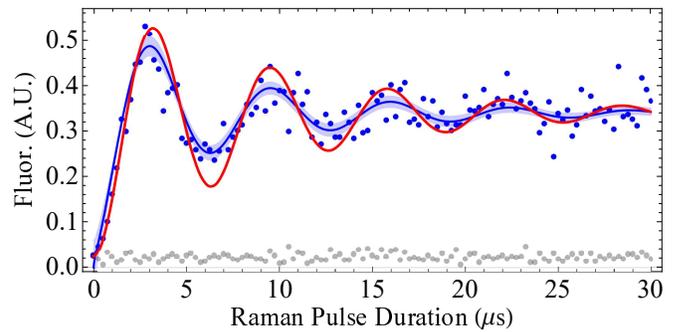}
\caption{Nuclear state-selective Raman driving of the electronic spin.  After nuclear polarization into $m_I=+1$, a Raman pulse of variable duration is applied on the $\pN\rightarrow\pN$ transition (blue), after which the electronic population in $\p$ is measured.  The grey data represent a control experiment during which no Raman pulse was applied, and the red plot is a simulation of the Raman dynamics, as described in the Supplemental Materials \cite{SuppInfo}.}
\label{fig.Electronic Rabi}
\end{center}
\end{figure}

Further, we demonstrate that we can drive coherent oscillations of the electronic spin that are conditioned on the nuclear spin state, as was first observed in Ref. \cite{Golter2014a}.  After initializing the electronic spin to $\z$ and the nuclear spin to $\pN$, we apply a Raman pulse of variable duration before reading out the population that has been transferred to $\p$.  As shown in Fig. \ref{fig.Electronic Rabi}, we observe coherent oscillations of the electronic spin between $\z$ and $\p$ as we vary the length of the Raman pulse.

\begin{figure}
\begin{center}
\includegraphics[width=\columnwidth]{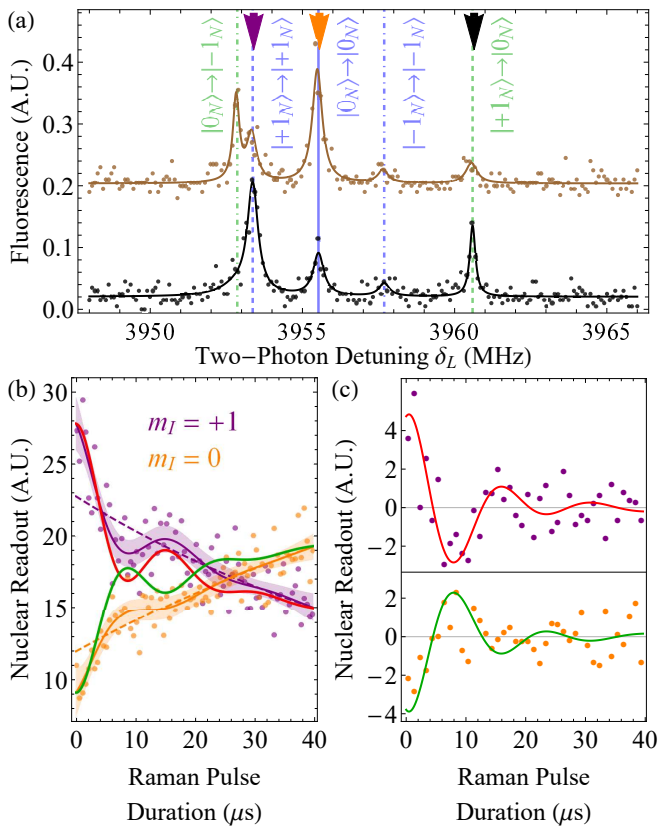}
\caption{Optical driving of the nuclear spin. (a) Spectroscopy of the Raman transition, conducted after a 40 $\mu$s pulse applied on the $\pN\rightarrow\zN$ transition (brown, offset vertically for clarity) and after an equivalent wait time (black).  (b) The $\pN$ and $\zN$ populations measured as a function of the duration of the $\pN\rightarrow\zN$ Raman pulse, applied at the frequency indicated by the black arrow in (a).  The purple and orange plots are fits to the data with 95\% mean prediction interval and the green and red plots are the results of a simulation of the Raman dynamics, as described in the Supplemental Materials \cite{SuppInfo}.  In (c), we subtract off the exponential decay component of the data and simulations shown in (b) \cite{SuppInfo}.}
\label{fig.Nuclear Rabi}
\end{center}
\end{figure}

Finally, we demonstrate coherent manipulation of  an electronic-nuclear flip-flop transition.  To this end, we first show  that we can transfer population from $\pN$ to $\zN$ by driving the $\pN\rightarrow\zN$ Raman transition.  We use a technique that is similar to that used to demonstrate polarization of the nuclear spin in \crefformat{figure}{Fig.~#2#1{(b)}#3}\cref{fig.Nuclear Polarization} \cite{SuppInfo}.  We polarize the nuclear spin to $\pN$ and the electronic spin to $\z$ and then either drive the $\pN\rightarrow\zN$ transition or wait for an equivalent length of time before performing spectroscopy of the Raman transitions.  The results of the spectroscopy, which are shown in \crefformat{figure}{Fig.~#2#1{(a)}#3}\cref{fig.Nuclear Rabi}, indicate that applying the Raman pulse on the $\pN\rightarrow\zN$ transition indeed transfers population from $\pN$ to $\zN$.  Without the $\pN\rightarrow\zN$ Raman pulse (black plot), the peaks that correspond to Raman transitions originating in $\pN$ are more prominent, but when the $\pN\rightarrow\zN$ Raman pulse is applied (brown plot), the peaks that correspond to Raman transitions originating in $\zN$ are more prominent and those that correspond to Raman transitions originating in $\pN$ are suppressed.  This indicates that we can transfer nuclear population optically.

To probe the dynamics of nuclear population transfer, we vary the length of the $\pN\rightarrow\zN$ Raman pulse and measure the subsequent populations of $\pN$ and $\zN$.  To read out the nuclear spin state, we map it to the electronic spin state by applying a $\pi$ pulse on one of the three nuclear spin-conserving transitions and then read out the electronic spin population optically.  Because the optical transition we use to read out the $\p$ population also pumps efficiently to $\z$, this nuclear readout process can be repeated many times in order to increase the information acquired during each run.  We perform this experiment twice, first applying the readout $\pi$ pulse on the $\pN\rightarrow\pN$ transition [purple arrow in \crefformat{figure}{Fig.~#2#1{(a)}#3}\cref{fig.Nuclear Rabi}] to read out a signal proportional to the $\pN$ population [purple plot in \crefformat{figure}{Fig.~#2#1{(b)}#3}\cref{fig.Nuclear Rabi}] and then similarly measuring the $\zN$ population (orange arrow and plot).  We note that, with optimization of the optical transitions that are used to read out and pump the electronic spin, this technique could potentially enable all-optical single-shot readout of the nuclear spin.

The results of this measurement are shown in \crefformat{figure}{Fig.~#2#1{(b)}#3}\cref{fig.Nuclear Rabi}.  While the dominant feature is incoherent population transfer from $\pN$ to $\zN$, there also appears to be evidence of coherent oscillations.  We fit the nuclear population data in \crefformat{figure}{Fig.~#2#1{(b)}#3}\cref{fig.Nuclear Rabi} to the sum of an exponential ramp and an exponentially damped sinusoid.  We also simulate the expected nuclear Raman dynamics for both the nuclear spin-conserving transition (red plot in Fig. \ref{fig.Electronic Rabi}) and the electronic-nuclear flip-flop transition (red and green plots in Fig. \ref{fig.Nuclear Rabi}), assuming that decoherence of the Rabi oscillations is due solely to off-resonant optical excitation \cite{SuppInfo}.  To match the data, we treat the Raman laser power and slight misalignment of the magnetic field as free parameters that are common to all three plots.  We set the simulations' vertical offsets to match the initial values of the data fits in \crefformat{figure}{Fig.~#2#1{(b)}#3}\cref{fig.Nuclear Rabi} or the mean of the grey reference measurement in Fig. \ref{fig.Electronic Rabi}, and we set their vertical scalings to match the final values of the respective data fits.

 \crefformat{figure}{Figure~#2#1{(c)}#3}\cref{fig.Nuclear Rabi} shows the extracted  sinusoidal components of the data and simulation plots shown in \crefformat{figure}{Fig.~#2#1{(b)}#3}\cref{fig.Nuclear Rabi}.  This is done by fitting each to the sum of an exponential ramp and an exponentially damped sinusoid, as in \crefformat{figure}{Fig.~#2#1{(b)}#3}\cref{fig.Nuclear Rabi}, and then subtracting off the exponential ramp and constant background terms.  The oscillations in the data and simulations have similar periods in all four cases but have opposite phases depending on the state that is being read out, as expected.  Thus, our observations constitute the onset of
coherent nuclear dynamics. However, the coherent oscillations are  suppressed by decoherence due to optical pumping at a rate that is comparable to the Raman Rabi frequency.

To understand these observations and to explore the options for improving the nuclear optical control, we present a detailed theoretical analysis of this coupled electron-nuclear system \cite{SuppInfo}.  While in general one can increase the Raman detuning $\Delta$ in order to suppress the incoherent optical pumping rate $\Gamma$ relative to the Raman Rabi frequency $\tilde{\Omega}$, this approach is limited by destructive interference between the Raman transitions driven via $\Eone$ and $\Ey$ states, as we calculate explicitly in the Supplemental Materials \cite{SuppInfo}.  Ultimately, the ratio $\tilde{\Omega}/\Gamma$ is determined by the strength of the interaction that mediates that specific Raman transition relative to the total decay rate out of the intermediate state.  Therefore, while the coherence of the nuclear-spin conserving transitions may be improved by applying a small magnetic field transverse to the N-V axis to couple $\Eone$ and $\Ey$ via the Zeeman interaction, the coherence of the electronic-nuclear flip-flop transitions is limited by the fact that the transverse hyperfine coupling rate $\lambda$ is comparable to the radiative decay rate $\gamma$ out of the $^3E$ states ($\lambda/\gamma \approx 2$).

In conclusion, we have experimentally demonstrated control of a single nuclear spin state via coherent optical excitation.  These proof-of-concept experiments can be extended along several promising directions.  Better control of the transverse magnetic field would enable a precise measurement of the transverse hyperfine coupling strength in the excited state and a more detailed investigation of the ESLAC-based nuclear polarization mechanisms.  Furthermore, similar techniques could be readily applied to other solid-state defects, such as silicon-vacancy centers, in which all-optical coherent manipulation of the SiV center's electronic spin has already been demonstrated \cite{Rogers2014c,Pingault2014a}, or the divacancy defect center in 3C-SiC, which has a similar excited state structure and radiative decay rate but a hyperfine coupling rate of approximately 50 MHz with a proximal $^{13}$C nucleus \cite{Christle2017}.  This optical manipulation technique could be particularly useful for densely spaced arrays of color centers integrated into nanophotonic devices \cite{Evans2018}, where it would enable the proximal nuclear spins to serve as the memory ancilla that are crucial for building quantum networks.

\begin{acknowledgments}
The authors would like to thank S. Singh and S. Choi for stimulating discussions. This work was supported by NSF, CUA, the DARPA QUEST program, AFOSR MURI, ARC, Element Six, and the Packard Foundation.  This material is based upon work supported by the National Science Foundation Graduate Research Fellowship under Grant No. DGE1745303.
\end{acknowledgments}

%

\end{document}


\newcommand{\figSize}{0.5}
\newcommand{\N}{^{14}\mathrm{N}}
\newcommand{\Hel}{H_\mathrm{el}}
\newcommand{\Ex}{|E_x\rangle}
\newcommand{\Ey}{|E_y\rangle}
\newcommand{\Eone}{|E_1\rangle}
\newcommand{\Etwo}{|E_2\rangle}
\newcommand{\Aone}{|A_1\rangle}
\newcommand{\Atwo}{|A_2\rangle}
\newcommand{\z}{|0\rangle}
\newcommand{\p}{|+1\rangle}
\newcommand{\m}{|-1\rangle}
\newcommand{\zN}{|0_N\rangle}
\newcommand{\pN}{|+1_N\rangle}
\newcommand{\mN}{|-1_N\rangle}
\newcommand{\At}{\,^3A_2}
\newcommand{\Et}{\,^3E}
\newcommand{\As}{|^1A_1\rangle}

\newcommand{\Pj}{|\Psi_j\rangle}
\newcommand{\Htot}{H_\mathrm{tot}}
\newcommand{\ExO}{|E_x^\prime\rangle}
\newcommand{\EyO}{|E_y^\prime\rangle}

\renewcommand{\thefigure}{S\arabic{figure}}
\renewcommand{\theequation}{S\arabic{equation}}

\title{Supplemental Material: Optical Control of a Single Nuclear Spin in the Solid State}

\author{M. L. Goldman}
\email[]{mgoldman@post.harvard.edu}
\affiliation{Department of Physics, Harvard University, Cambridge, Massachusetts 02138, USA}

\author{T. L. Patti}
\affiliation{Department of Physics, Harvard University, Cambridge, Massachusetts 02138, USA}

\author{D. Levonian}
\affiliation{Department of Physics, Harvard University, Cambridge, Massachusetts 02138, USA}

\author{S. F. Yelin}
\affiliation{Department of Physics, Harvard University, Cambridge, Massachusetts 02138, USA}
\affiliation{Department of Physics, University of Connecticut, Storrs, Connecticut 06269, USA}

\author{M. D. Lukin}
\affiliation{Department of Physics, Harvard University, Cambridge, Massachusetts 02138, USA}

\maketitle

\tableofcontents

\section{Experimental Details}
\label{sec.Experimental Details}

\subsection{NV Center Level Structure and ESLAC}
\label{sec.NV Center Level Structure and ESLAC}

\begin{figure}
\begin{center}
\begin{subfigure}[c]{0.17 \columnwidth}
\includegraphics[width=\columnwidth]{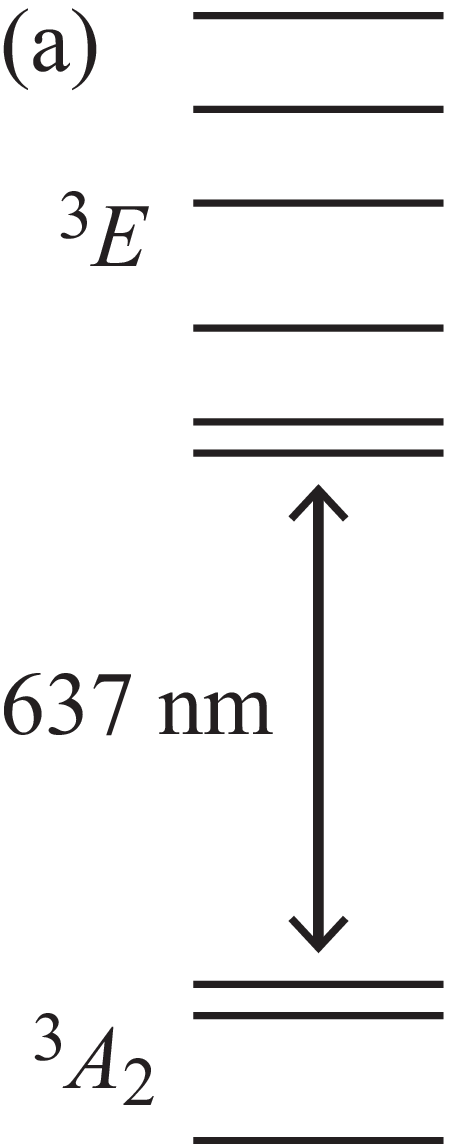}
\end{subfigure}
\hfill
\begin{subfigure}[c]{0.77 \columnwidth}
\includegraphics[width=\columnwidth]{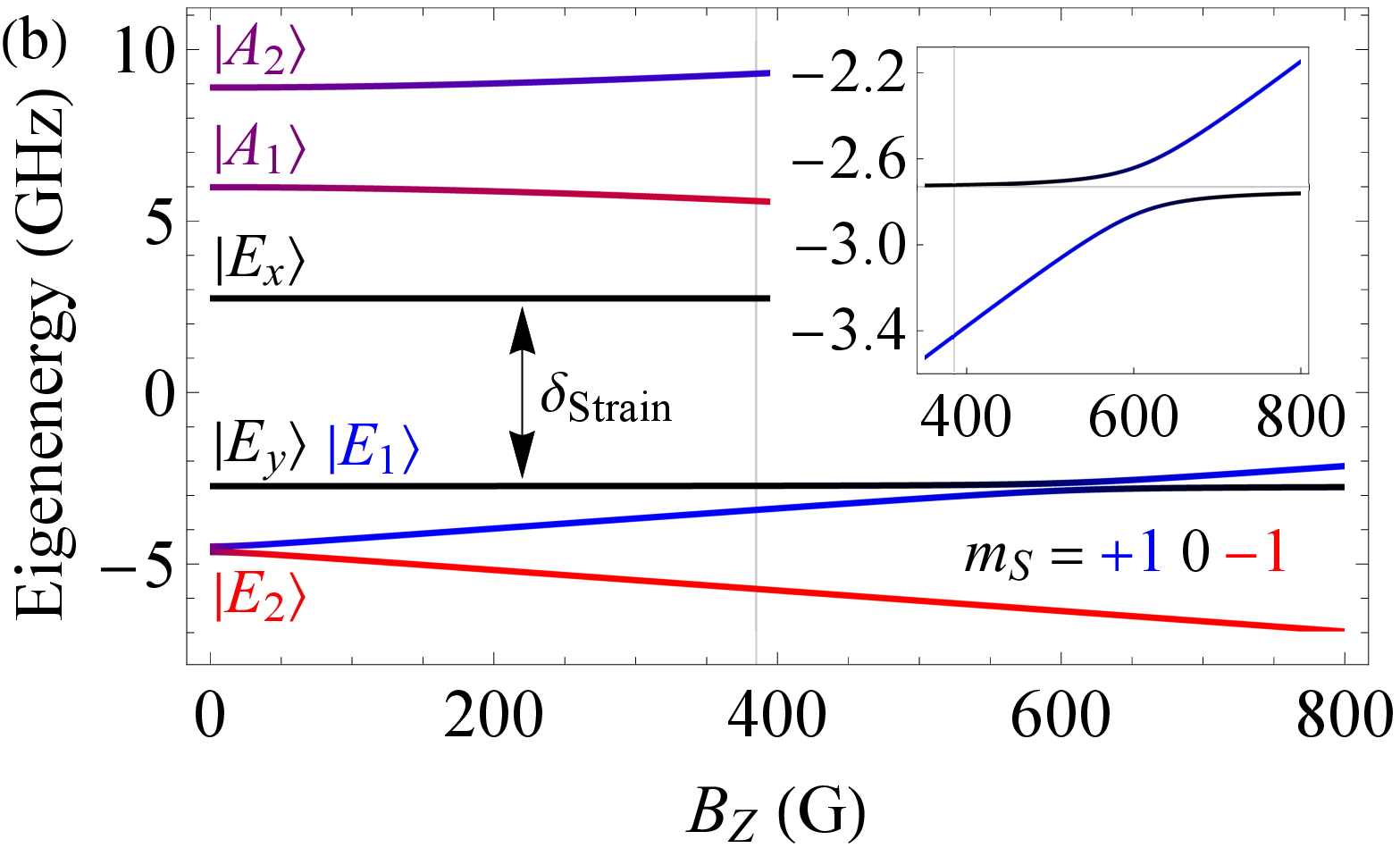}
\end{subfigure}
\caption{Level structure of the NV center.  (a) A schematic diagram of the NV center's spin-triplet, orbital-singlet ground state ($^3A_2$) and spin-triplet, orbital-doublet optically excited state ($^3E$).  (b) The calculated energies of the six states in the excited $^3E$ manifold as a function of the magnetic field $B_z$ applied along the N-V axis.  Each plot is colored according to that state's admixture of spin states with $m_S = +1$ (blue), 0 (black), or -1 (red).  The strain splitting between the $E_x$ and $E_y$ orbital branches is set to the value of 5.5 GHz measured for this NV center (see Sec. \ref{sec.Electronic Hamiltonian}).  The inset shows the anticrossing between $\Ey$ and $\Eone$, and the vertical lines in both plots indicate the applied field.}
\label{fig.Electronic Eigenstates Spectrum}
\end{center}
\end{figure}

The NV center has a spin-triplet, orbital-singlet ground state (labeled $^3A_2$) that is optically coupled to a spin-triplet, orbital-doublet excited state (labeled $^3E$), as shown in \crefformat{figure}{Fig.~#2#1{(a)}#3}\cref{fig.Electronic Eigenstates Spectrum}.  Strain in the diamond lattice splits the $^3E$ manifold into two orbital branches, as shown in \crefformat{figure}{Fig.~#2#1{(b)}#3}\cref{fig.Electronic Eigenstates Spectrum}.  A magnetic field applied along the N-V axis causes a Zeeman shift of the states with nonzero spin projection $m_S$, which results in an excited state level anticrossing (ESLAC) between the states in the lower orbital branch with primarily $m_S=0$ and $m_S=+1$ character, which are respectively labeled $\Ey$ and $\Eone$.  The specific interactions that give rise to the ESLAC are discussed in more detail in the main text.

\begin{figure}
\begin{center}
\includegraphics[width=\columnwidth]{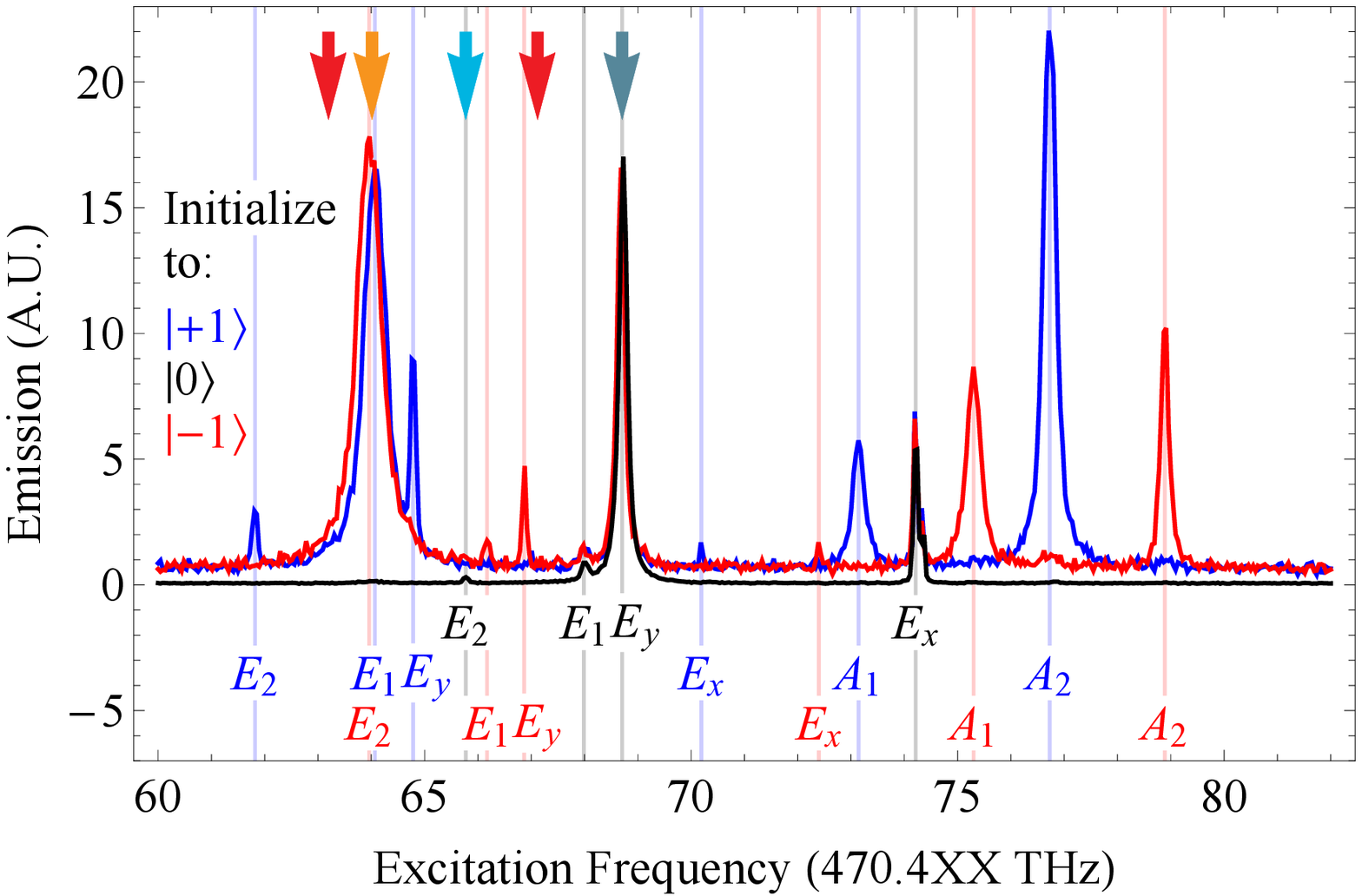}
\caption{Spectroscopy of the $\Et$ excited state manifold.  We measure the PLE spectra of the optical transitions between the $\At$ ground state triplet and $\Et$ excited state manifold.  The initial $\At$ state is indicated by the spectrum color and the final $\Et$ state is indicated below each line.  The arrows, whose colors match those in the pulse sequences shown in Fig. 2 and Secs. \ref{sec.Raman Spectroscopy Sequences} and \ref{sec.Raman Dynamics Sequences}, indicate the transitions or approximate frequencies used to (red) drive the off-resonant Raman transition, (orange) optically pump from $\p$ and $\m$ to $\z$, and (light or dark blue) optically pump from $\z$.}
\label{fig.Wavemeter Linescan}
\end{center}
\end{figure}

In order to identify and address the correct optical transitions, we use photoluminescence excitation (PLE) spectroscopy to probe the structure of the $^3E$ excited state manifold.  We initialize the NV center to $\z$ by applying green light at 520 nm and then pumping resonantly on the $|\pm1\rangle\rightarrow\Atwo$ transitions, or we initialize to $\p$ or $\m$ by initializing to $\z$ and then applying a non-hyperfine-selective microwave $\pi$ pulse.  We then probe the optical transitions to $^3E$ by applying a laser at 637 nm whose frequency is locked to a variable setpoint using a HighFinesse WS-6 wavelength meter.

The resulting spectra, which are shown in Fig. \ref{fig.Wavemeter Linescan}, indicate that we can resolve 16 of the 18 possible optical transitions.  We can identify the specific transitions by comparing these spectra to the calculated eigenstates of the $\Et$ manifold, as described in Sec. \ref{sec.Electronic Hamiltonian}.  We then use the wavelength meter to lock our lasers to the appropriate frequencies for driving the Raman and pumping transitions indicated by the arrows in Fig. \ref{fig.Wavemeter Linescan}.

\subsection{Raman Spectroscopy Sequences}
\label{sec.Raman Spectroscopy Sequences}

\begin{figure}[h]
\begin{center}
\includegraphics[width=\columnwidth]{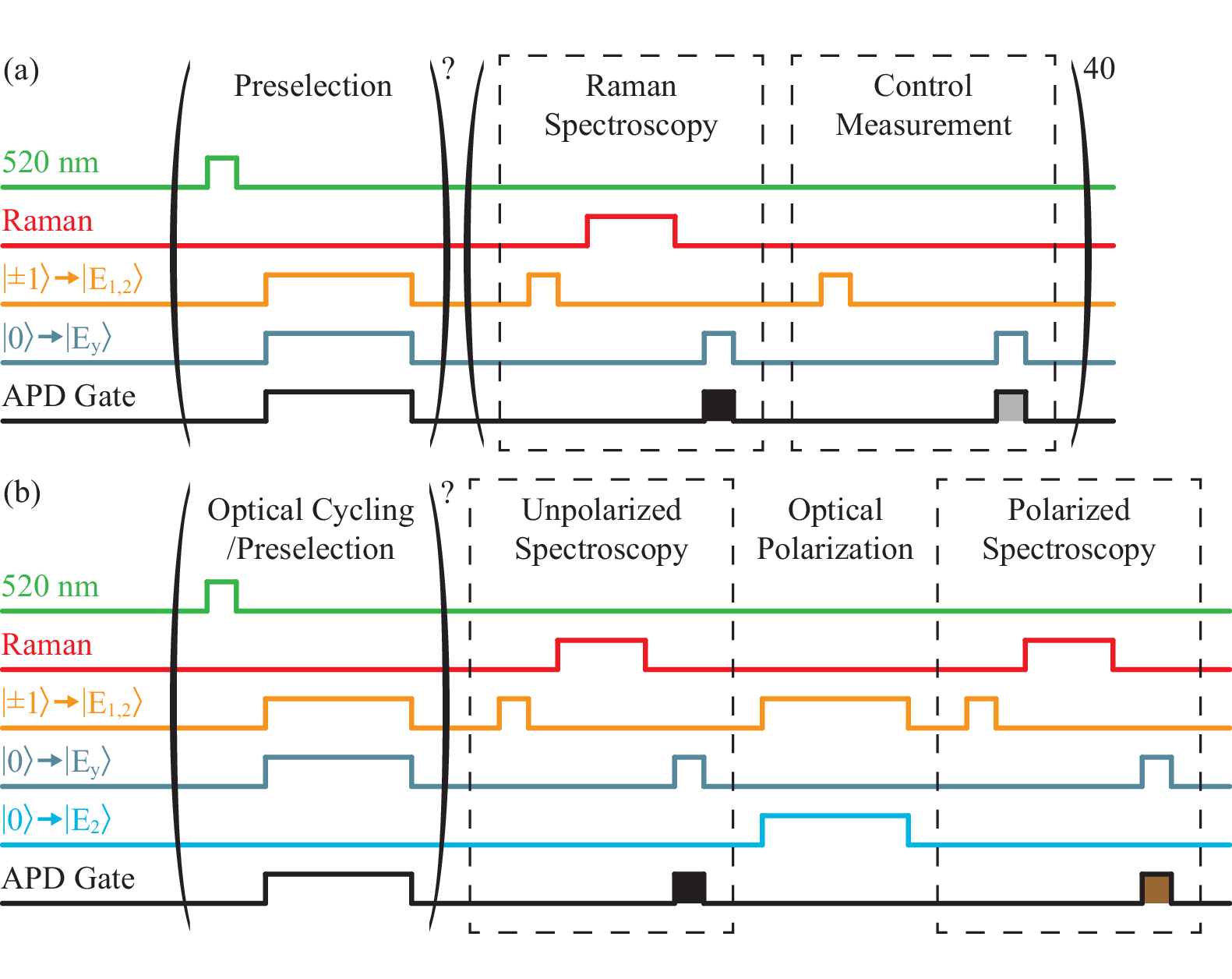}
\end{center}
\caption{Schematic illustrations of the pulse sequences used to acquire the (a) unpolarized and (b) polarized Raman spectroscopy data shown, respectively, in Figs. 1 and 2.  The colors of the APD Gate windows match those of the corresponding data sets.}
\label{fig.Raman ESR Pulse Sequence}
\end{figure}

The Raman spectroscopy measurement is the basic unit with which our various measurements are built.  In Fig. \ref{fig.Raman ESR Pulse Sequence}, we illustrate the pulse sequence used to acquire the Raman spectroscopy data shown in Figs. 1 and 2 in the main text.

We use the pulse sequence illustrated in \crefformat{figure}{Fig.~#2#1{(a)}#3}\cref{fig.Raman ESR Pulse Sequence} to acquire the unpolarized Raman spectroscopy data shown in Fig. 1(c).  First, we apply light at 520 nm to initialize the NV center to the negatively charged state, which causes spectral diffusion of the NV center's optical transitions \cite{Bassett2011}. We negate this spectral diffusion with a preselection stage that tests whether the NV center is in the correct charge state and whether its transitions are resonant with the excitation lasers \cite{Robledo2010}, which are applied for 120 $\mu$s on transitions from all three ground states.  This preselection stage is repeated until a sufficient number of photons are detected during the test window.

Next, we perform two Raman spectroscopy sequences.  In each, we first polarize the electronic spin to $\z$ by pumping on the $\p\rightarrow\Eone$ and $\m\rightarrow\Etwo$ transitions for 4 $\mu$s.  These transitions are strong because $\Eone$ and $\Etwo$ are mostly polarized into $m_S=+1$ and $m_S=-1$, respectively, by the axial magnetic field and these transitions pump efficiently to $\z$ because of the significant admixture of $m_S=0$ character in $\Eone$.  Moreover, these transitions are nearly degenerate (see Fig. \ref{fig.Wavemeter Linescan}) because the electronic $g$-factors in the ground and excited states are similar, so they can both be driven efficiently by a single laser.  We next apply the two Raman lasers for 15 $\mu$s before reading out the population that remains in $\z$ by exciting the $\z\rightarrow\Ey$ transition for 4 $\mu$s.

We next perform a control measurement by repeating the Raman spectroscopy sequence without the Raman pulse.  We create the two sidebands that drive the Raman transition by applying a modulation frequency $f_\mathrm{mod}=\delta_L/2$ to an electro-optic amplitude modulator (EOM), where $\delta_L$ is the Raman two-photon detuning.  We then repeat this pair of sequences 40 times per preselection cycle and sweep $f_\mathrm{mod}$ to map out the range of $\delta_L$ shown in Fig. 1(c).

We use a similar sequence, which is illustrated in \crefformat{figure}{Fig.~#2#1{(b)}#3}\cref{fig.Raman ESR Pulse Sequence}, to characterize the nuclear polarization mechanism, as shown in 2(b).  As with the Raman spectroscopy measurement, the preselection is performed by pumping on optical transitions from all three ground states that do not, in aggregate, result in polarization of the nuclear spin. For this measurement, the green laser was applied for 12 $\mu$s and the preselection lasers were applied for 30 $\mu$s.  The green-preselection stage is collectively represented by the ``Optical Cycling'' stage shown in 2(a) in the main text.

After the preselection, we perform two identical copies of the Raman spectroscopy sequence described above with a nuclear polarization stage in between, using a Raman pulse length of 8 $\mu$s for this measurement.  The nuclear polarization stage consists of optically pumping from $\p$ and $\m$ on the same transitions as in the preselection stage and pumping from $\z$ on the transition to $\Etwo$ instead of $\Ey$, as we discuss in Sec. \ref{sec.Polarization Mechanism}.  This sequence is only performed once per green-preselection cycle.  We again sweep the EOM modulation frequency $f_\mathrm{mod}$ to probe all of the possible Raman transitions.

\subsection{Raman Dynamics Sequences}
\label{sec.Raman Dynamics Sequences}

\begin{figure}[ht!]
\begin{center}
\includegraphics[width=\columnwidth]{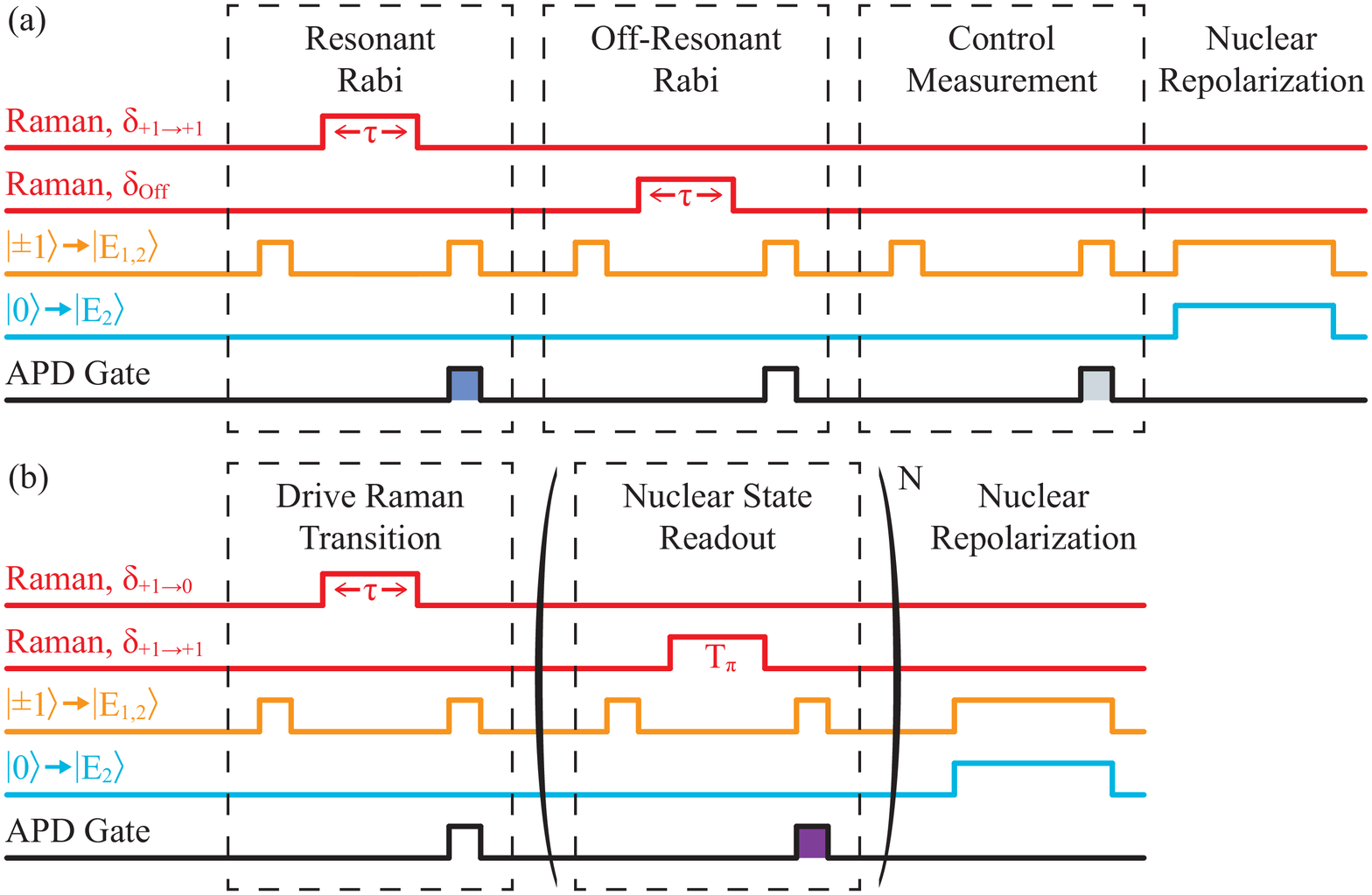}
\end{center}
\caption{Schematic illustration of the pulse sequences used to measure population dynamics (a) on the  $\pN\rightarrow\pN$ transition, as shown in Fig. 3, and (b) on the $\pN\rightarrow\zN$ transition, as shown in Fig. 4.  The green-preselection stage is omitted for concision.}
\label{fig.Raman Rabi Pulse Sequence}
\end{figure}

We use the sequences illustrated in \ref{fig.Raman Rabi Pulse Sequence} to measure population dynamics on the $\pN\rightarrow\pN$ and $\pN\rightarrow\zN$ transitions.  For simplicity, we do not show the green-preselection stage.  Unlike in Fig. \ref{fig.Raman ESR Pulse Sequence}, we use the $\z\rightarrow\Etwo$ transition for preselection instead of the $\z\rightarrow\Ey$ transition so that we can polarize the nuclear spin during the preselection stage.  As in \crefformat{figure}{Fig.~#2#1{(a)}#3}\cref{fig.Raman ESR Pulse Sequence}, we perform the sequences shown in Fig. \ref{fig.Raman Rabi Pulse Sequence} 40 times per green-preselection cycle.

We use the sequence illustrated in \crefformat{figure}{Fig.~#2#1{(a)}#3}\cref{fig.Raman Rabi Pulse Sequence} to measure population dynamics on the $\pN\rightarrow\pN$ transition, as shown in Fig. 3.  In this sequence, we perform three iterations of the Raman spectroscopy sequence before repolarizing the nuclear spin to $\pN$ by pumping on the nuclear-polarizing transitions for 20 $\mu$s.  In all iterations of the spectroscopy sequence, we excite the $\p\rightarrow\Eone$ and $\m\rightarrow\Etwo$ transitions to read out the population that has been transferred to $\p$.  We tune $\delta_L$ to the $\pN\rightarrow\pN$ transition for the first iteration, we tune $\delta_L$ 2.7 MHz below the $\pN\rightarrow\pN$ transition for the second iteration, and we omit the Raman pulse for the third iteration.  We sweep the duration of both Raman pulses to map out the population dynamics during Raman driving.

We use the sequence illustrated in \crefformat{figure}{Fig.~#2#1{(b)}#3}\cref{fig.Raman Rabi Pulse Sequence} to measure population dynamics on the $\pN\rightarrow\zN$ transition, as shown in Fig. 4(b).  In this measurement, we first perform a Raman spectroscopy sequence with $\delta_L$ tuned to the $\pN\rightarrow\zN$ transition and a variable Raman pulse duration.  We next perform the Raman spectroscopy sequence many times with $\delta_L$ tuned to either the $\pN\rightarrow\pN$ or $\zN\rightarrow\zN$ transition.  This process repetitively maps the nuclear spin population to the electronic spin so that it can be read out.  The length of this Raman pulse is set to 3.21 $\mu$s, which was the measured value that maximized the population transfer from $\z$ to $\p$.  This mapping sequence is performed 10 times with the $\pN\rightarrow\pN$ transition and 30 times with the $\zN\rightarrow\zN$ transition, and in Fig. 4 we plot the average signal per repetition.  After these repetitions, we optically pump for 20 $\mu$s to repolarize the nuclear spin.

The sequence we use to measure nuclear population transfer, as shown in Fig. 4(a), is similar to that illustrated in \crefformat{figure}{Fig.~#2#1{(b)}#3}\cref{fig.Raman Rabi Pulse Sequence} but with three differences.  First, we fix the length of the Raman pulse applied on the $\pN\rightarrow\zN$ transition to 40 $\mu$s for the brown plot, and we replace that pulse with 40 $\mu$s of waiting for the black plot.  In both cases, we remove the brief readout pulse on the $\p\rightarrow\Eone$ and $\m\rightarrow\Etwo$ transitions.  Second, we only perform one subsequent Raman spectroscopy sequence to read out the nuclear spin.  Third, we fix the length of the nuclear readout Raman pulse to 15 $\mu$s and we sweep the EOM modulation frequency $f_\mathrm{mod}$ in order to probe all of the Raman transitions.

\subsection{Characterization of Phonon-Induced Mixing}
\label{sec.Characterization of Phonon-Induced Mixing}

In this section, we characterize the rate of phononic mixing between the $\Et$ states, which contributes to the incoherent optical pumping that limits the coherence of our Raman manipulation, as discussed in Sec. \ref{sec.Incoherent Optical Pumping Rates}.  We measure this mixing rate using the same technique used in Ref. \cite{Goldman2015} to measure the intersystem crossing (ISC) rates from the individual $\Et$ states.  We apply a short optical pulse to excite the NV center to the $\Atwo$ state and then measure the decay timescale of the resulting fluorescence.  Because the ISC rate from $\Atwo$ is negligible and we apply only a small magnetic field for this measurement, which should not appreciably mix $\Atwo$ and $\Aone$, we would expect the decay timescale to be roughly equal to the radiative decay timescale measured in Ref. \cite{Goldman2015}.

Using the same analysis technique as in Ref. \cite{Goldman2015}, we extract an effective ISC rate of $\tilde{\Gamma}_{A_2}/2\pi=1.2\pm0.5$ MHz.  We convert this ISC rate to a sample temperature using the fit and 95\% confidence interval of the ISC rate from $\Atwo$ shown in Fig. 3 of Ref. \cite{Goldman2015}.  We find a sample temperature of $14\pm1$ K, which implies that sample mounting chip used to apply our microwave pulses raises the sample temperature by approximately 7 K relative to the coldfinger temperature.  We emphasize that microwave pulses were only used for characterization measurements and were not used in any of the pulse sequences described in the previous two sections.

The coldfinger temperature was set 1.7 K higher for the Raman measurements shown in Fig. 4 than for this thermometry measurement, so we add the same offset to the extracted sample temperature.  We then use the fit of the phonon-induced mixing rate as a function of temperature, which is also described in Ref. \cite{Goldman2015}, to convert this sample temperature to a mixing rate of $2\pi\times(4.1\pm1.7)$ MHz.

\section{Nuclear Polarization}
\label{sec.Nuclear Polarization}

\subsection{Polarization Mechanism}
\label{sec.Polarization Mechanism}

We now present a qualitative explanation of the nuclear polarization mechanism.  The two spectra shown in Fig. 2(b) demonstrate that the transition we use to repump the NV center from $\z$ to $\p$ or $\m$ determines whether the nuclear spin is polarized.  We can understand this behavior by considering the mechanisms that enable the electronic spin to flip under optical illumination and by analyzing how these mechanisms can also flip the nuclear spin.

We first consider the case, labeled ``Optical Cycling'' in Fig. 2(a), of optical pumping that does not polarize the nuclear spin.  We excite from $\z$ to $\Ey$ and from $\p$ to $\Eone$, which are both strong transitions because $\Ey$ and $\Eone$ respectively have mostly $m_S=0$ and $m_S=+1$ character [see Figs. \ref{fig.Wavemeter Linescan} and \crefformat{figure}{#2#1{(b)}#3}\cref{fig.Electronic Eigenstates}].  Therefore, after excitation, the NV center will most likely decay back to the original ground state.  However, because we are operating in the vicinity of the ESLAC, where $\Ey$ and $\Eone$ are substantially mixed, there is a significant possibility that the NV center will flip the electronic spin by decaying from $\Ey$ to $\p$ ($\Delta m_S=+1$) or from $\Eone$ to $\z$ ($\Delta m_S=-1$).  Because $\Ey$ and $\Eone$ are mixed, in part, by the transverse nuclear hyperfine interaction, there is some probability that flipping the electronic spin by $\Delta m_S=+1$ or $-1$ will be accompanied, respectively, by flipping the nuclear spin by $\Delta m_I=-1$ or $+1$; this is precisely the same physics that gives rise to the Raman transitions that conserve the total spin by flipping both the electronic and nuclear spins.  Because of the symmetry of these pumping dynamics --- we pump the electronic spin back and forth between $\z$ and $\p$ by passing through the same ESLAC in both directions --- no net nuclear polarization is induced.

We expect that pumping to and from $\m$ has a relatively small effect on nuclear polarization.  We repump from $\m$ via $\Etwo$, which has mostly $m_S=-1$ character.  There are nonzero branching ratios from both $\Ey$ and $\Eone$ to $\m$ but they are relatively small, as are the branching ratios from $\Etwo$ to $\p$ and $\z$.  In fact, the dominant decay paths to and from $\m$ are expected to be via the intersystem crossing (ISC) through the spin-singlet states.  In the spin-singlet states, the two unpaired electrons occupy the same single-electron orbital states as they do in the spin-triplet $\At$ ground state, which have relatively little spatial overlap with the $\N$ nucleus.  It is therefore reasonable to assume that the hyperfine interaction in the spin-singlet states is relatively weak, as it is in the $\At$ ground state, and that the ISC decay process therefore has a negligible effect on the nuclear spin.  Thus, the processes of decaying to and repumping from $\m$ likely have little effect on the nuclear polarization in the case of optical cycling.

We now consider the case of optical polarization of the nuclear spin.  In this case, we repump from $\z$ by exciting not the strong $\z\rightarrow\Ey$ transition but rather the weak $\z\rightarrow\Etwo$ transition.  $\Etwo$ is coupled to $\Ex$ by the spin-spin, transverse Zeeman, and transverse hyperfine interactions, just as $\Eone$ is coupled to $\Ey$.  However, $\Etwo$ and $\Ex$ are separated, for the NV center studied in this Letter, by more than 8 GHz, which is much larger than the $\sim700$ MHz separation between $\Eone$ and $\Ey$.  The admixture of $m_S=0$ character in $\Etwo$ is accordingly very small, which explains why the $\z\rightarrow\Etwo$ transition is barely visible in the PLE spectrum shown in Fig. \ref{fig.Wavemeter Linescan}.  Nevertheless, the transition is allowed and $\Etwo$, once populated, decays with high probability into $\m$, flipping the electronic spin by $\Delta m_S=-1$.  From $\m$, we excite the NV center back to $\Etwo$, from which it eventually decays back to $\p$ and $\z$ via the ISC.  The key to the nuclear polarization mechanism is that pumping from $\p$ to $\z$ via $\Eone$ and pumping from $\z$ to $\m$ via $\Etwo$ both involve using the transverse hyperfine interaction to flip the electronic spin by $\Delta m_S=-1$, which is accompanied by flipping the nuclear spin by $\Delta m_I=+1$.  In the steady state, therefore, this optical cycling results in a net polarization into $\pN$.

\subsection{Comparison with Nonresonant Polarization Mechanism}
\label{sec.Comparison with Nonresonant Polarization Mechanism}

\begin{figure}
\begin{center}
\includegraphics[width=\columnwidth]{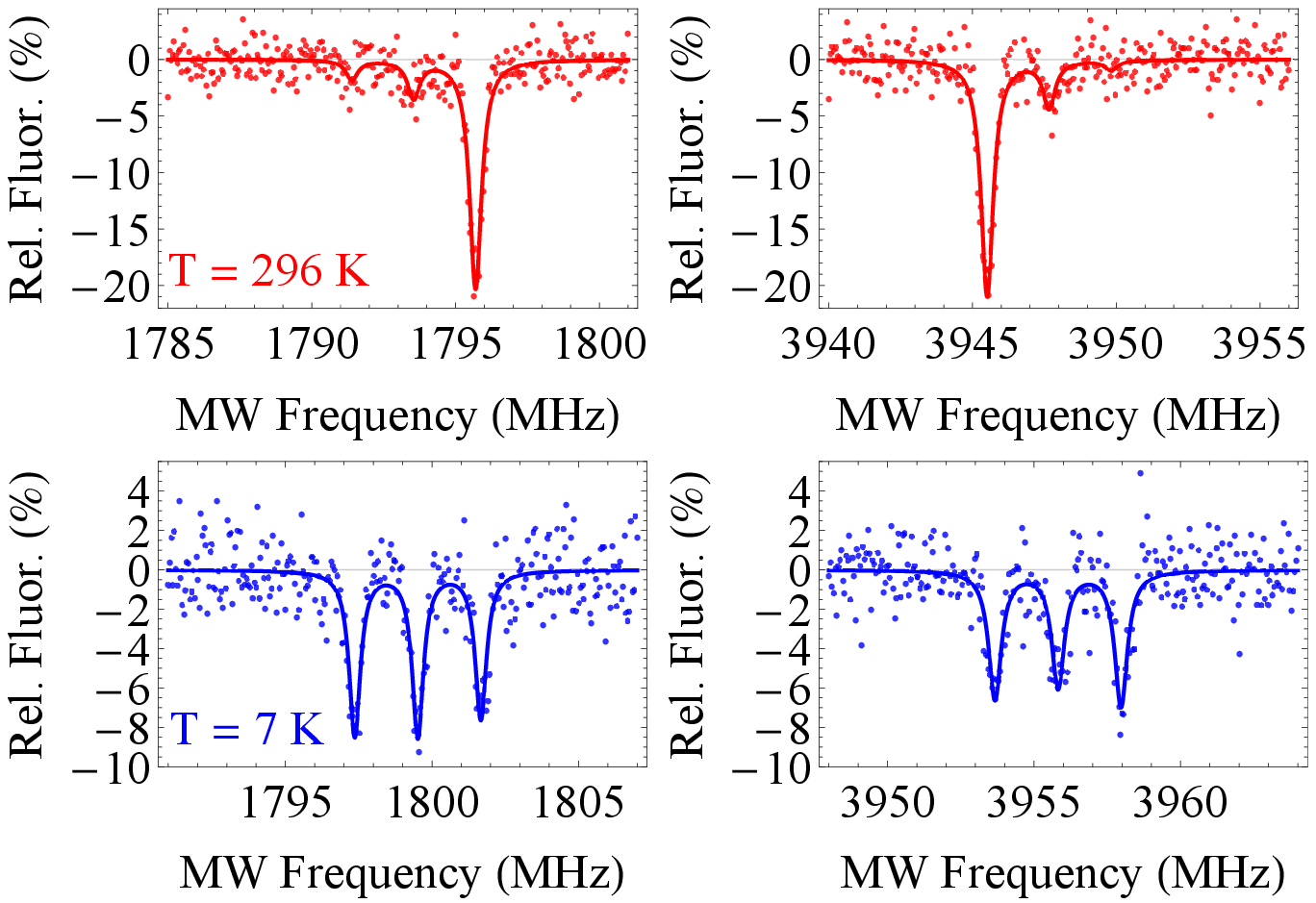}
\caption{ODMR spectra of the $\z\rightarrow\m$ (left) and $\z\rightarrow\p$ (right) transitions, measured at temperatures of 296 K (top) and 7 K (bottom).  The presence of three dips is due to hyperfine coupling with the $\N$ nucleus and imbalance between the depths of the three dips indicates polarization of the $\N$ spin.  All four measurements were conducted using nonresonant optical initialization and readout of the electronic spin at 520 nm.}
\label{fig.ESR Comparison}
\end{center}
\end{figure}

We claim that this optical $\N$ spin polarization mechanism is distinct from the nonresonant mechanism that has previously been studied at room temperature, which also uses the ESLAC to transfer polarization from the electronic spin to the nuclear spin \cite{Jacques2009,Smeltzer2009,Steiner2010,Poggiali2017}.  We do, in fact, observe polarization due to this nonresonant mechanism at room temperature but not at cryogenic temperatures.  We use microwave fields to perform optically detected magnetic resonance (ODMR) spectroscopy of the $\z\rightarrow\m$ and $\z\rightarrow\p$ transitions, which is the same technique used in Refs. \cite{Jacques2009,Smeltzer2009,Steiner2010,Poggiali2017} to demonstrate nuclear polarization, and display the results in Fig. \ref{fig.ESR Comparison}.  Spectroscopy of the $\z\rightarrow\p$ and $\z\rightarrow\m$ transitions, respectively, indicate 81(3)\% and 80(3)\% polarization into the $\pN$ state, but only at room temperature.  This degree of nuclear polarization is to be expected, since the polarization mechanism is effective over a range of several hundred gauss around 500 G \cite{Jacques2009} and our magnetic field was permanently aligned at room temperature to be parallel to the N-V axis.

We note that all four measurements shown in Fig. \ref{fig.ESR Comparison} were conducted using the same pulse sequence, with the only differences, beyond the sample temperature, being the range of microwave frequencies scanned and the nominal microwave power applied.  We adjusted the nominal microwave power in order to compensate for frequency-dependent attenuation between the microwave source and the NV center, so that the $\pi$ pulse duration was approximately constant for all measurements [1.35 (1.40) $\mu$s for the $\z\rightarrow\p(\m)$ transition].

The fact that we do not observe significant nuclear polarization due to the nonresonant mechanism at cryogenic temperatures, even when such polarization is present at room temperature, indicates that this mechanism is not responsible for the polarization observed in Fig. 2.  There are two possible reasons that we do not observe this nonresonant mechanism at lower temperatures.  First, the degree of polarization caused by the nonresonant mechanism is sensitive to a field misalignment of $\sim1^{\circ}$.  Because the magnetic field was aligned at room temperature to be parallel to the N-V axis, cooling the cryostat from 296 K to 7 K may have caused enough mechanical drift due to differential thermal expansion to cause significant misalignment of the magnetic field.  The ODMR spectra shown in Fig. \ref{fig.ESR Comparison} indicate that the axial field shifts by approximately 400 mG, suggesting the magnetic field does change as the sample is cooled down, and the numerical simulation of the Raman dynamics described in Sec. \ref{sec.Simulation of Coherent Dynamics} is consistent with a magnetic field misalignment of approximately $5^{\circ}$.

Alternatively, orbital averaging within the $\Et$ manifold, which is suppressed at cryogenic temperatures \cite{Goldman2015}, may be necessary to enable the off-resonant polarization mechanism.  The polarization mechanism has been treated theoretically in the presence of orbital averaging \cite{Ivady2015,Nah2016,Poggiali2017}, which enables the $\Et$ manifold to be treated more simply as two orbital branches that each consist of a spin triplet with an effective zero-field splitting of 1.43 GHz \cite{Fuchs2008}.  In this picture, orbital averaging essentially elides details of the $\Et$ fine structure and sample-to-sample inhomogeneities, such as crystal strain \cite{Rogers2009}.  However, evaluating how this polarization mechanism functions in the cryogenic regime, where the phonon-induced mixing between the two orbital branches is slower than the optical decay rate \cite{Goldman2015}, would require a detailed treatment of the $\Et$ fine structure.

To better understand this polarization mechanism, one could carefully measure the degree of nuclear polarization induced by both the resonant and nonresonant mechanisms as a function of magnetic field misalignment and temperature.  Combining this experimental input with the detailed theoretical description of the $\Et$ manifold given in Sec. \ref{sec.Theoretical Model of Raman Manipulation} and the model of phononic mixing described in Ref. \cite{Goldman2015a} would enable the development of a unified theoretical framework that describes how the $\N$ nuclear spin can be polarized under a range of experimental conditions.

\section{Theoretical Model of Raman Manipulation}
\label{sec.Theoretical Model of Raman Manipulation}

\subsection{Toy Raman Model}
\label{sec.Toy Raman Model}

\begin{figure}
\begin{center}
\includegraphics[width=\columnwidth]{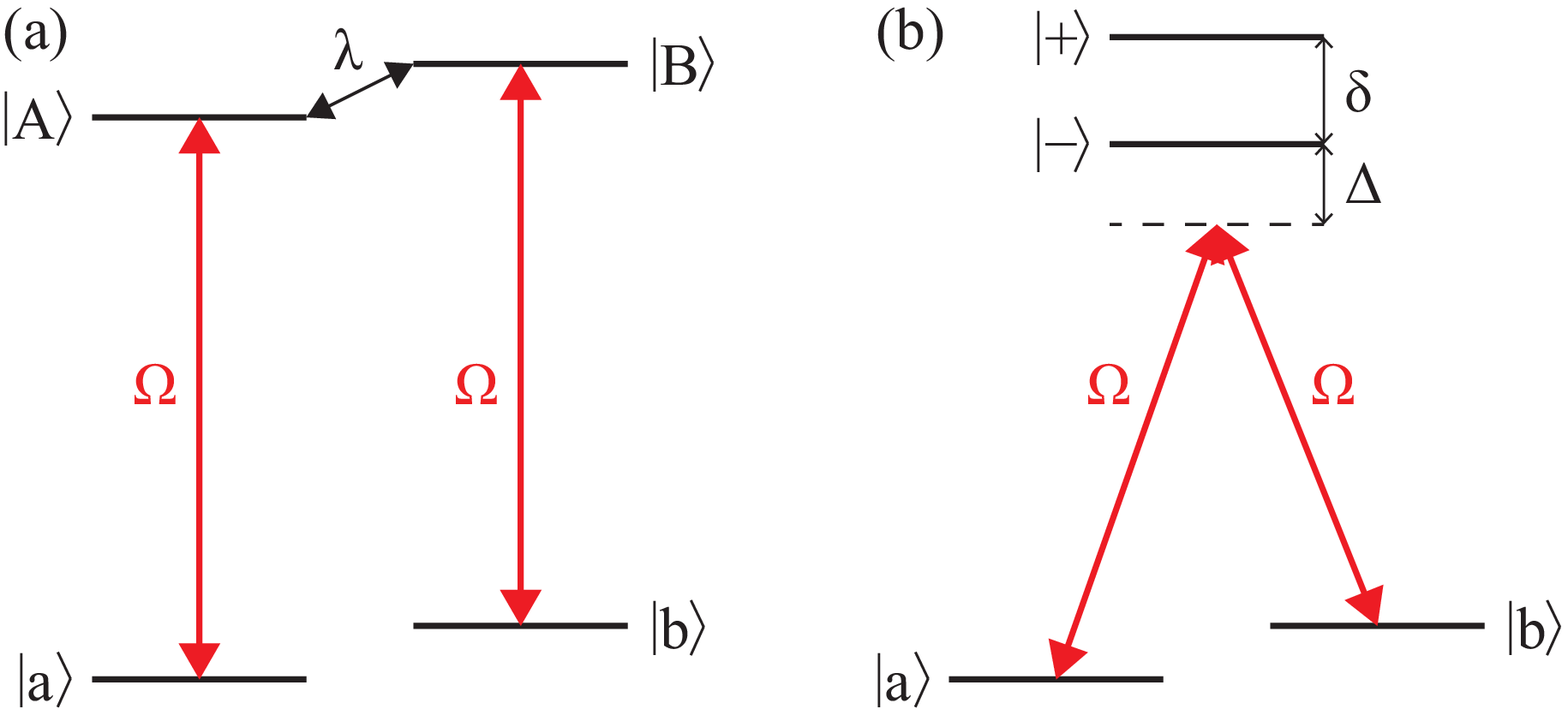}
\caption{Level diagram of the Raman toy model, drawn with the (a) bare excited states and (b) excited states dressed by the interaction of strength $\lambda$.}
\label{fig.Toy Model}
\end{center}
\end{figure}

We now develop a theoretical model that provides insight into the factors that limit the coherence of the Raman manipulation techniques described in the main text.  Our model includes all of the states of the excited $\Et$ manifold and uses the experimentally measured strengths of all of the terms of the $\Et$ Hamiltonian to calculate the precise electronic and nuclear spin admixtures of each state.  We use this information to calculate the strengths of the coherent Raman couplings between different ground states as well as the off-resonant optical pumping rates that we believe limit the coherence of our Raman manipulation.  We then estimate how we might optimize our experimental parameters in order to maximize the strength of the coherent Raman coupling relative to the incoherent pumping rates.

First, we may develop an intuition for the inherent limitations of our Raman manipulation by considering a simple toy model.  As illustrated in Fig. \ref{fig.Toy Model}, we consider two ground states $|a\rangle$ and $|b\rangle$ that are optically coupled to the excited states $|A\rangle$ and $|B\rangle$ with a common optical Rabi frequency $\Omega$.  The states $|A\rangle$ and $|B\rangle$, which are not necessarily degenerate, are coupled by an interaction of strength $\lambda$, so we may define the dressed states
\begin{align}
\label{eq.Raman dressed states}
|+\rangle &= \alpha |A\rangle + \beta |B\rangle \nonumber \\
|-\rangle &= \beta |A\rangle - \alpha |B\rangle,
\end{align}
where we take $\alpha$ and $\beta$ to be real.  These dressed states are separated in energy by $\delta$ and we detune the two driving lasers by $\Delta$ below the transitions from $|a\rangle$ and $|b\rangle$ to $|-\rangle$.  These two lasers can drive a Raman transition between $|a\rangle$ and $|b\rangle$ with a Rabi frequency
\begin{equation}
\label{eq.Raman Rabi freq 1}
\tilde{\Omega}_{ab} = \tilde{\Omega}_{ab}^{(+)} + \tilde{\Omega}_{ab}^{(-)} = \frac{\Omega_{b+}^\ast \Omega_{a+}}{\Delta_+} + \frac{\Omega_{b-}^\ast \Omega_{a-}}{\Delta_-},
\end{equation}
where $\tilde{\Omega}_{ab}^{(+)}$ ($\tilde{\Omega}_{ab}^{(-)}$) is the Rabi frequency of the Raman transition driven specifically via the dressed state $|+\rangle$ ($|-\rangle$); $\Omega_{a+}=\Omega\langle +|A\rangle$, etc., are the optical Rabi frequencies of the specific transitions between the given ground and dressed excited states; and $\Delta_+=\Delta+\delta$ ($\Delta_-=\Delta$) is the one-photon detuning from $|+\rangle$ ($|-\rangle$).  Because $|+\rangle$ and $|-\rangle$ are orthogonal, the relative minus sign in Eq. \ref{eq.Raman dressed states} necessarily appears in the definition of one of the states and so the Raman transitions driven through $|+\rangle$ and $|-\rangle$ interfere destructively with each other when the signs of $\Delta_+$ and $\Delta_-$ are the same.  In the far-detuned limit $|\Delta|\gg\delta$, the total Raman Rabi frequency becomes
\begin{align}
\tilde{\Omega}_{ab} &= \frac{|\Omega|^2 \langle B|+\rangle\langle +|A\rangle}{\Delta_+} + \frac{|\Omega|^2 \langle B|-\rangle\langle -|A\rangle}{\Delta_-} \nonumber \\
&= \frac{|\Omega|^2 \beta\alpha}{\Delta+\delta} + \frac{|\Omega|^2 (-\alpha)\beta}{\Delta} \nonumber \\
&\approx -\frac{|\Omega|^2 \alpha\beta}{\Delta}\frac{\delta}{\Delta}.
\end{align}

At the same time, these two lasers cause off-resonant optical pumping from both $|a\rangle$ and $|b\rangle$ at an average rate
\begin{align}
\Gamma_{ab} &= \frac{1}{2}\left(\Gamma_a + \Gamma_b\right) \nonumber \\
&= \frac{1}{2}\left(\frac{|\Omega_{a+}|^2\gamma}{\Delta_+^2} + \frac{|\Omega_{a-}|^2\gamma}{\Delta_-^2} + \frac{|\Omega_{b+}|^2\gamma}{\Delta_+^2} + \frac{|\Omega_{b-}|^2\gamma}{\Delta_-^2} \right) \nonumber \\
&\approx \frac{|\Omega|^2\gamma}{\Delta^2},
\end{align}
where $\Gamma_a$ ($\Gamma_b$) is the optical pumping rate out of $|a\rangle$ ($|b\rangle$) and $\gamma$ is the common radiative decay rate from $|A\rangle$ and $|B\rangle$.  Because both the Raman Rabi frequency and the optical pumping rate scale as $|\Omega|^2/\Delta^2$ in the far-detuned limit, the advantage that can be gained by simply applying more laser power and detuning farther from the optical transitions saturates quickly once $\Delta>\delta$.

More explicitly, the figure of merit
\begin{equation}
\label{eq.Raman figure of merit}
\frac{|\tilde{\Omega}_{ab}|}{\Gamma_{ab}} = \alpha\beta \frac{\delta}{\gamma}
\end{equation}
has a very simple form and equally simple interpretation: we want the dressed states $|+\rangle$ and $|-\rangle$ to be roughly equal superpositions of $|A\rangle$ and $|B\rangle$ (i.e., maximize the product $\alpha\beta$) in order to enable strong optical coupling to both ground states, but we also want the dressed states $|+\rangle$ and $|-\rangle$ to be widely separated (i.e., $\delta\gg\gamma$) in order to minimize the effect of destructive interference between the Raman transitions driven via the two dressed states.

We might ask whether we can maximize $|\tilde{\Omega}|/\Gamma$ by either tuning $|A\rangle$ and $|B\rangle$ to degeneracy to maximize the degree of mixing or by making them widely separated in order to decrease the destructive interference.  Neither approach presents an obvious advantage.  In the case of degeneracy, the dressed states $|+\rangle$ and $|-\rangle$ are split by twice the coupling strength ($\delta=2\lambda$) and are equal superpositions of $|A\rangle$ and $|B\rangle$ ($\alpha=\beta=1/\sqrt{2}$), so we find that Eq. \ref{eq.Raman figure of merit} simply reduces to
\begin{equation}
\label{eq.Raman figure of merit 2}
\frac{|\tilde{\Omega}_{ab}|}{\Gamma_{ab}} = \frac{\lambda}{\gamma}.
\end{equation}
In the widely detuned case, when we separate $|A\rangle$ and $|B\rangle$ by a large applied splitting $\Delta_\mathrm{app}\gg\lambda$, we find that $|+\rangle \approx |B\rangle + \lambda/\Delta_\mathrm{app} |A\rangle$ and $\delta \approx \Delta_\mathrm{app}$, which is also results in the simple ratio given by Eq. \ref{eq.Raman figure of merit 2}.  We note also that tuning the lasers between $|+\rangle$ and $|-\rangle$ (i.e. $\Delta_+ = -\Delta_- = -\delta/2$) so that the Raman transitions through the two dressed states interfere constructively also results in Eqs. \ref{eq.Raman figure of merit} and \ref{eq.Raman figure of merit 2}.

Fundamentally, then, the degree to which we can drive a Raman transition coherently is simply given by the ratio of the strength $\lambda$ of the interaction that enables the Raman transition to the decay rate $\gamma$ that enables off-resonant optical pumping, the primary decoherence mechanism.  This result implies that the ratio $|\tilde{\Omega}|/\Gamma_{ab}$ is, to a large degree, set by the fundamental physics of the system and cannot be drastically changed by varying experimental parameters.  We next check the results of this toy model by considering the details of the interactions that give rise to the actual Raman transitions that are the focus of this work.

\subsection{Electronic Hamiltonian}
\label{sec.Electronic Hamiltonian}

\begin{figure}
\begin{center}
\includegraphics[width=\columnwidth]{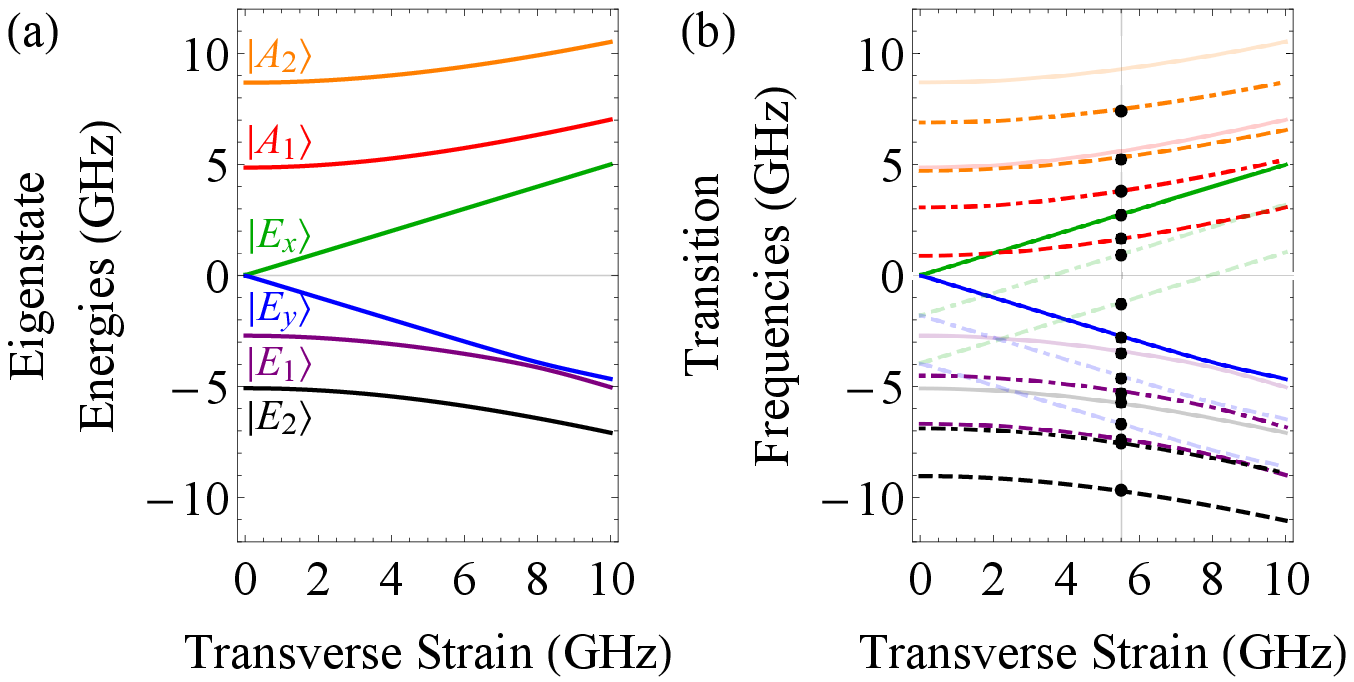}
\caption{Diagonalization of the $\Et$ electronic Hamiltonian.  We construct and numerically diagonalize the electronic Hamiltonian $H_\mathrm{el}$, as described in the text.  In (a), the eigenenergies of the six $\Et$ states are calculated as a function of the transverse strain splitting $\delta$.  In (b), we compare the calculated transition energies from the $|0\rangle$ (solid), $|+1\rangle$ (dashed), and $|-1\rangle$ (dotdashed) ground states to the measured transition frequencies extracted from the PLE spectra shown in Fig. \ref{fig.Wavemeter Linescan} (black dots).  The low-opacity plots in (b) represent relatively weak transitions.  The theoretical transition frequencies are fitted to the measured transition frequencies using only the strain splitting and average transition frequency as free parameters.}
\label{fig.Electronic Eigenstates}
\end{center}
\end{figure}

In order to build our model, we first construct the electronic Hamiltonian that determines the fine structure of the $\Et$ manifold.  This Hamiltonian is given by
\begin{equation}
\label{eq.Hel}
H_\mathrm{el} = H_\mathrm{SO} + H_\mathrm{SS} + H_\mathrm{Strain} + H_\mathrm{Z,s} + H_\mathrm{Z,l},
\end{equation}
which includes terms describing the electrons' spin-orbit interaction ($H_\mathrm{SO}$), spin-spin interaction ($H_\mathrm{SS}$), interaction with crystal strain ($H_\mathrm{Strain}$), and Zeeman interaction of a magnetic field with the electronic spin ($H_\mathrm{Z,s}$) and orbital angular momentum ($H_\mathrm{Z,l}$).  We use the explicit forms of each of these Hamiltonian terms derived by Doherty, et al. \cite{Doherty2011} but we use the state labels adopted by Maze, et al. \cite{Maze2011}, which are common in the experimental NV center literature.  Explicitly, our basis states, which are eigenstates of the spin-orbit Hamiltonian that dominates at zero crystal strain and applied magnetic field, are
\begin{align}
\label{eq.Basis States}
|A_1^{(0)}\rangle &= \frac{1}{\sqrt{2}}\big( |+1\rangle \otimes |E_-^\prime\rangle + |-1\rangle \otimes |E_+^\prime\rangle \big) \nonumber \\
|A_2^{(0)}\rangle &= \frac{-i}{\sqrt{2}}\big( |+1\rangle \otimes |E_-^\prime\rangle - |-1\rangle \otimes |E_+^\prime\rangle \big) \nonumber \\
|E_x^{(0)}\rangle &= |0\rangle \otimes \ExO \nonumber \\
|E_y^{(0)}\rangle &= |0\rangle \otimes \EyO \nonumber \\
|E_1^{(0)}\rangle &= \frac{i}{\sqrt{2}}\big( |+1\rangle \otimes |E_+^\prime\rangle - |-1\rangle \otimes |E_-^\prime\rangle \big) \nonumber \\
|E_2^{(0)}\rangle &= \frac{1}{\sqrt{2}}\big( |+1\rangle \otimes |E_+^\prime\rangle + |-1\rangle \otimes |E_-^\prime\rangle \big),
\end{align}
where $\p$, $\m$, and $\z$ are the electronic spin wavefunctions, $\ExO$ and $\EyO$ are the electronic orbital states with zero angular momentum, and
\begin{equation}
|E_\pm^\prime\rangle = \mp \frac{1}{\sqrt{2}}\big( \ExO \pm i \EyO \big)
\end{equation}
are the electronic orbital states with nonzero angular momentum.  We include the prime symbol, which is not standard notation, in order to differentiate explicitly between the orbital states $\ExO$ and $\EyO$ and the eigenstates $\Ex$ and $\Ey$ of the full electronic Hamiltonian.

One consideration that warrants discussion is the effect of coupling between the magnetic field and the electrons' orbital angular momentum, which Doherty, et al. do not express explicitly in terms of a measured parameter.  Following Bassett, et al. \cite{Bassett2014}, we define this interaction as
\begin{equation}
\label{eq.HZl1}
H_\mathrm{Z,l} = \frac{1}{2} \, \mu_B \, g_\mathrm{orb} \, B_z \, \sigma_y \otimes I_3,
\end{equation}
where $\mu_B$ is the Bohr magneton, $g_\mathrm{orb} =2\,L_z/\mu_B$ parameterizes the $z$ component of the orbital angular momentum, $B_z$ is the axial magnetic field (relative to the N-V axis), $\sigma_y$ is a Pauli matrix that operates in the $\{\ExO,\EyO\}$ orbital basis, and $I_3$ is an identity matrix that operates in the spin basis.  There are no interactions between the orbital angular momentum and transverse magnetic fields \cite{Bassett2014}.  Translating into the basis given by Eq. \ref{eq.Basis States}, we find
\begin{align}
\label{eq.HZl2}
H_\mathrm{Z,l} = \frac{i}{2}\, \mu_B \, g_\mathrm{orb} \, B_z  \large( & |A_1^{(0)}\rangle\langle A_2^{(0)}| + |E_y^{(0)}\rangle\langle E_x^{(0)}| \nonumber \\
& + |E_2^{(0)}\rangle\langle E_1^{(0)}| \large) + \mathrm{h.c}.
\end{align}

We use the values for the spin-orbit and spin-spin interaction strengths measured by ultrafast optical spectroscopy \cite{Bassett2014} and the experimentally measured values for the ground and excited state axial $g$-factors $g^{||}_\mathrm{GS} = 2.0028 (3)$ \cite{Loubser1978,He1993,Felton2009}, $g^{||}_\mathrm{ES} = 2.15 (4)$ \cite{Bassett2014}, $g^{||}_\mathrm{orb} = 0.10(1)$ \cite{Reddy1987,Rogers2009}.  We use the value of the axial magnetic field measured in Sec. \ref{sec.Comparison with Nonresonant Polarization Mechanism} and we assume negligible off-axis field.  The crystal strain is a free parameter and is quantized in terms of the energy splitting $\delta$ between $\Ex$ and $\Ey$.

We diagonalize $H_\mathrm{el}$ to give the energies of the six $^3E$ eigenstates, which are plotted in \crefformat{figure}{Fig.~#2#1{(a)}#3}\cref{fig.Electronic Eigenstates} as a function of the crystal strain $\delta$.  We then use the measured separations of the $\At$ states (see Sec. \ref{sec.Comparison with Nonresonant Polarization Mechanism}) to calculate the relative frequencies of all optical transitions from the ground state triplet to the six excited states, which are plotted in \crefformat{figure}{Fig.~#2#1{(b)}#3}\cref{fig.Electronic Eigenstates}.  We compare these calculated transition frequencies to those that we extract from the PLE spectrum shown in Fig. \ref{fig.Wavemeter Linescan}.  We minimize the sum of the absolute values of the errors between the measured and calculated transition frequencies in order to extract a strain splitting of $\delta = 5.5$ GHz.  We find excellent agreement between the measured and calculated relative transition frequencies.

\subsection{Hyperfine Hamiltonian}
\label{sec.Hyperfine Hamiltonian}

We now extend our model to include the nuclear degree of freedom.  The electronic-nuclear Hamiltonian in the excited state is given by
\begin{equation}
H_\mathrm{nuc} = A_{||}^\mathrm{ES}S_z I_z + A_\perp^\mathrm{ES}\left(S_+ I_- + S_- I_+\right) + P I_z^2 + \gamma_{\N}\mathbf{B}\cdot\mathbf{I},
\end{equation}
where $A_{||}^\mathrm{ES}$ and $A_\perp^\mathrm{ES}$ are, respectively, the axial and transverse hyperfine coupling rates, $P$ is the $\N$ quadrupolar shift, and $\gamma_{\N}=0.3077$ kHz/G is the $\N$ gyromagnetic ratio \cite{Felton2009,Smeltzer2009,Fuchs2011,Poggiali2017}.

The axial hyperfine coupling rate $A_{||}^\mathrm{ES}/h$ has been measured to be approximately 40 MHz \cite{Steiner2010}.  This is much stronger than the ground-state hyperfine coupling rate because optical excitation to the excited state results in a significant shift of unpaired electronic spin density toward the $\N$ nucleus, which significantly increases both the Fermi contact and dipolar interaction strengths \cite{Gali2009,Doherty2013a}.  Although the hyperfine interaction has been assumed to be isotropic \cite{Smeltzer2009,Steiner2010}, recent measurements have indicated that the transverse hyperfine rate is significantly smaller than the axial rate, with a value of $23\pm 3$ MHz \cite{Poggiali2017}.  Although there is disagreement in the literature on the sign of $A_{||}^\mathrm{ES}$ and $A_\perp^\mathrm{ES}$, physical reasoning about the mechanisms responsible for the hyperfine interaction \cite{Doherty2013a} and \textit{ab inito} calculations of the $^{15}$N hyperfine coupling parameters \cite{Gali2009} indicate that the signs of the hyperfine coupling rates in the ground and excited state should be different.  Therefore, since the ground-state hyperfine coupling rate has been established to be negative \cite{Felton2009,Smeltzer2009,Steiner2010}, we use $A_\perp^\mathrm{ES}/h=23$ MHz for the excited-state transverse hyperfine coupling rate.  Following Ref. \cite{Poggiali2017}, we assume that the quadrupolar shift $P$ has the same value in the excited state as in the ground state.

The total Hamilonian $\Htot$ is simply the sum of the electronic-nuclear Hamiltonian $H_\mathrm{nuc}$ and the electronic Hamiltonian $H_\mathrm{el}$.  We expand our basis from the 6 electronic states listed in Eq. \ref{eq.Basis States}, which are eigenstates of the spin-orbit Hamiltonian, to an 18-state basis, which is the tensor product of the electronic basis states with the nuclear basis states $\{\pN,\zN,\mN\}$.  Because the energy scales in $H_\mathrm{nuc}$ are generally smaller than those in $H_\mathrm{el}$, moving to this expanded basis essentially has the effect of splitting each of the electronic eigenstates whose energies are plotted in \crefformat{figure}{Fig.~#2#1{(a)}#3}\cref{fig.Electronic Eigenstates} into three electronic-nuclear eigenstates that are separated by a relatively small ($\lessapprox 80$ MHz) and electronic spin-dependent hyperfine splitting.

We can therefore express each eigenstate $\Pj$ of the total Hamiltonian $\Htot$ either as a tensor product of the six electronic eigenstates $\{\psi_i\}$ and three nuclear spin states $\{\pN,\zN,\mN\}$ or as a tensor product of the two electronic orbital states $O\in\{\ExO,\EyO\}$, three electronic spin states $\{\p,\z,\m\}$, and three nuclear spin states:
\begin{align}
\Pj &= \sum_{i,m_I} c_{i,m_I}^{(j)} |\psi_i,m_I\rangle \nonumber \\
&= \sum_{O\in\{E_x^\prime,E_y^\prime\},m_S,m_I} c_{O,m_S,m_I}^{(j)} |O,m_S,m_I\rangle.
\end{align}
The electronic and nuclear spin makeup of each eigenstate $\Pj$, which is characterized by the coefficient $c_{O,m_S,m_I}^{(j)}$, directly determines how that eigenstate optically couples to the various ground spin states.

\subsection{Coherent Raman Rabi Frequencies}
\label{sec.Coherent Raman Rabi Frequencies}

We calculate the effective Rabi frequencies of the various Raman transitions between the $\z$ and $\p$ ground states that we consider in the main text.  We consider a transition between the ground states $|a\rangle=|m_S=0,m_I=m_I^{(a)}\rangle$ and $|b\rangle=|m_S=+1,m_I=m_I^{(b)}\rangle$ driven via the eigenstate $\Pj$.

The ground states $|a\rangle$ and $|b\rangle$ are optically coupled to the excited eigenstate $\Pj$ with Rabi frequencies
\begin{align}
\label{eq.Optical Rabi frequencies}
\Omega_{a,j} &= \Omega \langle \Psi_j| E_y^\prime,m_S=0,m_I=m_I^{(a)}\rangle =\Omega \,  c_{E_y^\prime,0,m_I^{(a)}}^{(j)\ast} \nonumber \\
\Omega_{b,j} &= \Omega \langle \Psi_j| E_y^\prime,m_S=+1,m_I=m_I^{(b)}\rangle =\Omega \, c_{E_y^\prime,+1,m_I^{(b)}}^{(j)\ast},
\end{align}
where we select the components of $\Pj$ with the $E_y^\prime$ orbital state because we address the two transitions with linearly polarized lasers whose polarizations are set parallel to the $E_y^\prime$ dipole axis.  The Raman Rabi frequency for the transition between $|a\rangle$ and $|b\rangle$ is therefore
\begin{equation}
\label{eq.Raman Rabi freq}
\tilde{\Omega}_{ab} = \sum\limits_j \tilde{\Omega}_{ab}^{(j)} = \sum\limits_j \frac{C_{ab}^{(j)}|\Omega|^2}{\Delta_j},
\end{equation}
where $\Delta_j$ is the one-photon detuning of the two driving lasers from $\Pj$ and we have defined the dimensionless transition strength
\begin{equation}
\label{eq.Transition strength}
C_{ab}^{(j)} = c_{E_y^\prime,+1,m_I^{(b)}}^{(j)}c_{E_y^\prime,0,m_I^{(a)}}^{(j)\ast}.
\end{equation}

\begin{figure}
\begin{center}
\includegraphics[width=\columnwidth]{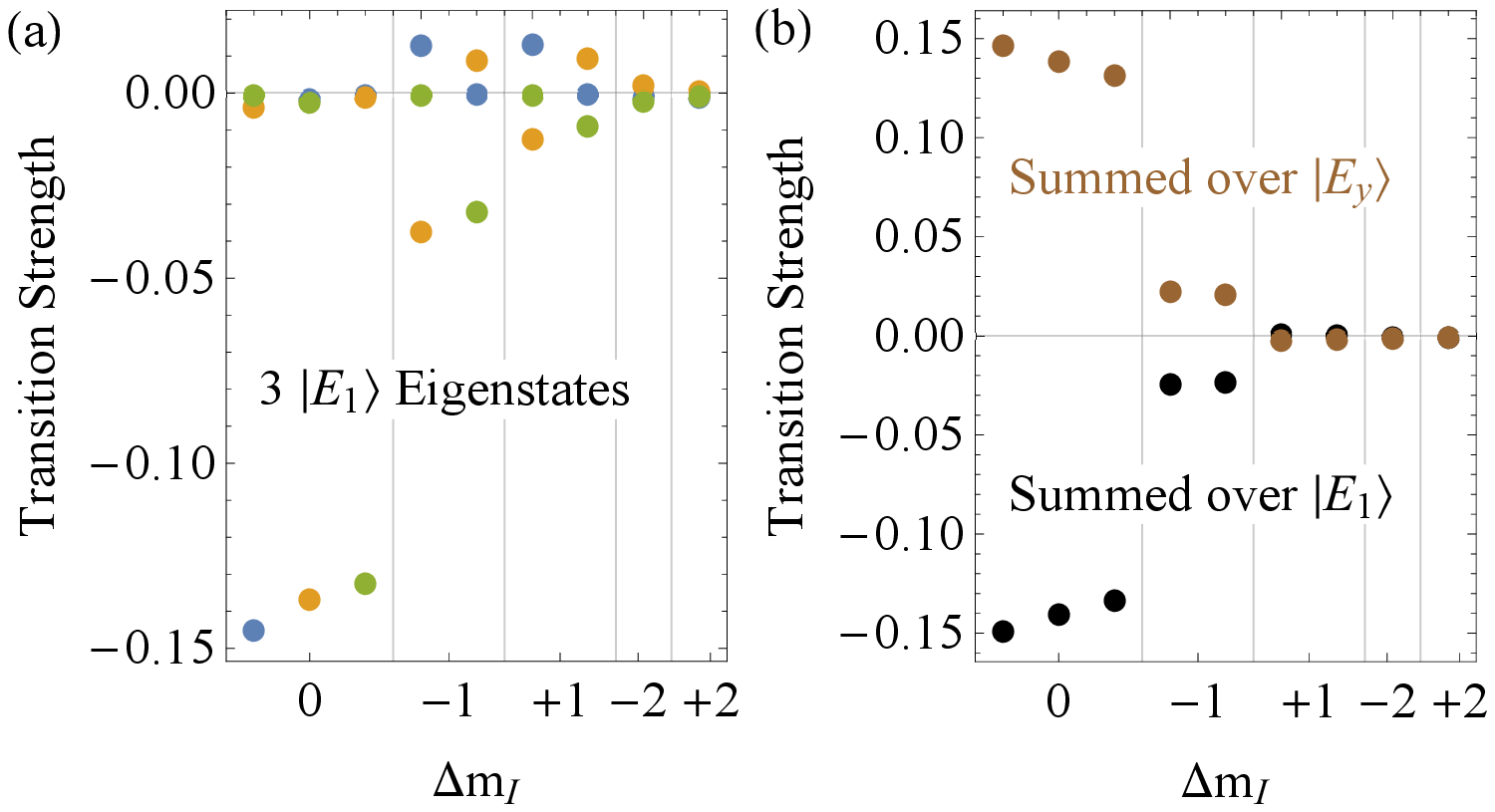}
\caption{The transition strengths $C_{ab}^{(j)}$ of the nine possible Raman transitions from $\z$ to $\p$ ($\Delta m_S=+1$), grouped by the change in $m_I$.  We plot (a) the transitions strengths for the three individual eigenstates corresponding to $\Eone$ and (b) the total transition strengths for $\Ey$ and $\Eone$, summed coherently over all three eigenstates.  Note that we do not consider the detunings $\Delta_j$ or laser power $|\Omega|^2$ in these plots.}
\label{fig.Raman transition strengths}
\end{center}
\end{figure}

To understand the observed Raman transition selection rules, we plot in Fig. \ref{fig.Raman transition strengths} the transition strengths $C_{ab}^{(j)}$ of the nine possible Raman transitions from $\z$ to $\p$ driven via the excited state $\Eone$.  We call attention to two features of these plots.  First, since $\Delta m_S=+1$ for this transition, we expect the transitions that conserve the nuclear spin ($\Delta m_I=0$) and the total spin ($\Delta m_I=-1$) to be allowed by the spin-spin and hyperfine interactions, respectively, and we expect all others to be suppressed.  This trend is generally observed when we consider the strengths of the Raman transitions driven through the individual electronic-nuclear eigenstates, as shown in Fig. \crefformat{figure}{~#2#1{(a)}#3}\cref{fig.Raman transition strengths}, but the distinction between the allowed and disallowed transitions is not especially stark.  When we sum the transition strengths over all three electronic-nuclear eigenstates, as shown in Fig. \crefformat{figure}{~#2#1{(b)}#3}\cref{fig.Raman transition strengths}, the expected hierarchy of transition strengths becomes immediately apparent.  As we would expect from our toy model, the ratio of the mean strength of the transitions with $\Delta m_I=0$ to the mean strength of the transitions with $\Delta m_I=-1$ (6.08) is approximately equal to the ratio of the interaction strengths that enable those transitions ($\lambda_\mathrm{ss}/A_\perp^\mathrm{ES}=6.70$).

Second, the summed strengths of the Raman transitions through the $\Eone$ and $\Ey$ eigenstates have nearly equal magnitudes but opposite signs.  This fact reflects the statement from the toy model that the Raman transitions driven through the two eigenstates $|+\rangle$ and $|-\rangle$ interfere destructively when $\Delta_+$ and $\Delta_-$ have the same sign, which is here generalized to destructive interference between the two groups of three eigenstates suppressing the rates of all nine Raman transitions.  Thus, the simple physical intuition that we extract from our toy model holds independently for the nuclear spin-conserving and total spin-conserving transitions, but only when coherence among the six relevant eigenstates is taken into account.  This coherent interference between the Raman transitions through $\Eone$ and $\Ey$ is precisely analogous to how the optically excited fine structure must be taken into account when calculating the rates of Raman processes in atomic species \cite{Ozeri2005}.

\subsection{Incoherent Optical Pumping Rates}
\label{sec.Incoherent Optical Pumping Rates}

We must compare the Raman Rabi frequencies to the rates of the processes that introduce decoherence into the Raman transition.  We assume that the dominant contribution to this decoherence is due to the fact that the lasers that drive the coherent Raman transition also incoherently excite population to the various excited eigenstates, which is the decoherence mechanism that we consider in our toy model.  The optical pumping rate from the ground state $|a\rangle$ to the excited eigenstate $\Pj$ is given by
\begin{equation}
\label{eq.Optical pumping rate}
\Gamma_a^{(j)} = \frac{|\Omega_{a,j}|^2}{\left(\gamma_j/2\right)^2+\Delta_j^2}\gamma_j = \frac{|\Omega|^2 \, C_{aa}^{(j)}}{\left(\gamma_j/2\right)^2+\Delta_j^2}\gamma_j,
\end{equation}
where $\gamma_j$ is the sum of the decay rates out of $\Pj$, $\Delta_j$ is the one-photon detuning of the pumping laser from the $|a\rangle\rightarrow\Pj$ transition, $\Omega$ and $\Omega_{a,j}$ are the bare and transition-specific optical Rabi frequencies defined in Eq. \ref{eq.Optical Rabi frequencies}, and $C_{aa}^{(j)}$ is the transition strength defined in Eq. \ref{eq.Transition strength}.

For completeness, we assume that the two applied Raman lasers can each off-resonantly pump from each of the nine ground states to each of the nine eigenstates corresponding to $\Ey$, $\Eone$, and $\Etwo$, which together comprise the entire lower orbital branch. We assign a unique decay rate $\gamma_j$ to each eigenstate $\Pj$ because $\gamma_j$ contains contributions not only from radiative decay, whose rate $\gamma$ is the same for all states in the $\Et$ manifold, but also from phonon-induced mixing between the various $\Et$ states and from the intersystem crossing (ISC) from $\Et$ states with $|m_S|=1$ character.  We estimate that the phonon-induced mixing, which we characterize in Sec. \ref{sec.Characterization of Phonon-Induced Mixing}, contributes approximately $2\pi\times4$ MHz to $\gamma_j$ for all eigenstates, which is small but not completely negligible compared to the radiative decay rate $\gamma=2\pi\times13$ MHz.  More careful thermal engineering could largely eliminate this contribution to $\gamma_j$ and could slightly reduce all pumping rates $\Gamma_a^{(j)}$.

The ISC provides an additional decay mechanism out of the $|A_1^{(0)}\rangle$, $|E_1^{(0)}\rangle$, and $|E_2^{(0)}\rangle$ states, which are approximately the electronic eigenstates in the regime of low magnetic field and crystal strain.  For these experiments, however, moderate crystal strain and the application of a substantial magnetic field have resulted in electronic eigenstates $\Pj$ that are superpositions of the electronic basis states $|A_1^{(0)}\rangle$, etc., along with an additional nuclear degree of freedom.  We can therefore define an effective ISC rate for each eigenstate $\Pj$, which is given by
\begin{align}
\Gamma_{\mathrm{ISC},j} =\, &|\langle A_1^{(0)}\Pj|^2\,\Gamma_{\mathrm{ISC},A_1} \nonumber \\
&+ \left(|\langle E_1^{(0)}\Pj|^2+|\langle E_2^{(0)}\Pj|^2\right)\Gamma_{\mathrm{ISC},E_{1,2}},
\end{align}
where $\Gamma_{\mathrm{ISC},A_1}$ and $\Gamma_{\mathrm{ISC},E_{1,2}}$ are the measured ISC rates from $|A_1^{(0)}\rangle$ and $|E_{1,2}^{(0)}\rangle$, respectively \cite{Goldman2015}.

We sum the different contributions to $\Gamma_{\mathrm{ISC},j}$ incoherently because the three ISC processes from $|A_1^{(0)}\rangle$, $|E_1^{(0)}\rangle$, and $|E_2^{(0)}\rangle$ result in distinguishable final states.  The ISC process from $|A_1^{(0)}\rangle$ does not involve the emission of an $E$-symmetric phonon, whereas the processes from $|E_1^{(0)}\rangle$ and $|E_2^{(0)}\rangle$ require phonons of different polarizations to couple the initial state to $|A_1^{(0)}\rangle$ \cite{Goldman2015a}.  As a result, the ISC process increases $\gamma_j$ by approximately $2\pi\times0.2$ MHz for the $\Ey$ eigenstates and by approximately $2\pi\times8.1$ MHz for the $\Eone$ eigenstates.

This increase in the incoherent pumping rates to the $\Eone$ eigenstates is mostly offset, however, by the fact that $\Eone$ has a smaller overlap with the $\EyO$ orbital state than $\Ey$, which reduces the factor $C_{aa}^{(j)}$ that enters into the incoherent pumping rate.  $\Eone$ has equal projections onto the $E_x^\prime$ and $E_y^\prime$ orbital states when the crystal strain splitting is low (cf. the electronic basis state $|E_1^{(0)}\rangle$ in Eq. \ref{eq.Basis States}), but its projection onto the $E_y^\prime$ orbital state increases as crystal strain splits the $\Et$ manifold into the $E_x^\prime$ and $E_y^\prime$ orbital branches.  Explicitly, we find that $|\langle E_y^\prime\Eone|^2\approx0.74$, which reduces the incoherent pumping rate to the $\Eone$ eigenstates by roughly the same factor that inclusion of the ISC decay increases those rates.

\subsection{Optimizing Parameters}
\label{sec.Optimizing Parameters}

To quantify the degree of coherence for a given Raman transition between states $|a\rangle$ and $|b\rangle$, we use the figure of merit $|\tilde{\Omega}_{ab}|/\Gamma_{ab}$ we defined in the context of the toy model, which is the ratio of the coherent Raman Rabi frequency to the average optical pumping rate out of the initial and final states.  This figure of merit generalizes easily to the full model, where $\tilde{\Omega}_{ab}$ is given by Eq. \ref{eq.Raman Rabi freq} and we define
\begin{equation}
\Gamma_{ab} = \frac{1}{2}\left(\Gamma_a+\Gamma_b\right) = \frac{1}{2}\sum\limits_j \left( \Gamma_a^{(j)}+\Gamma_b^{(j)}\right),
\end{equation}
where the pumping rates $\Gamma_a^{(j)}$ from a specific ground state to a specific excited eigenstate are given by Eq. \ref{eq.Optical pumping rate}.

\crefformat{figure}{Fig.~#2#1{(a)}#3}
\begin{figure}
\begin{center}
\includegraphics[width=\columnwidth]{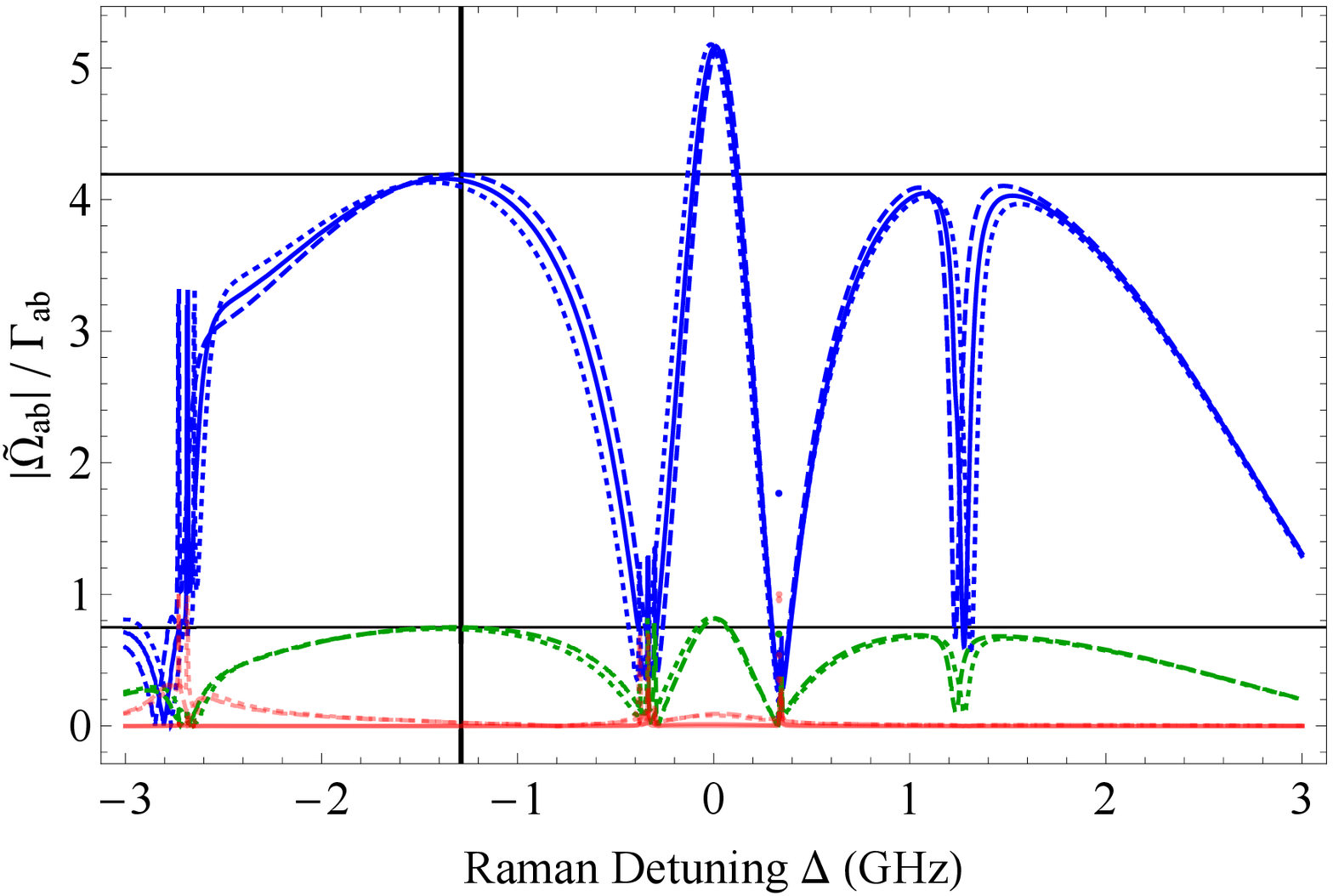}
\caption{The calculated ratio $|\tilde{\Omega}_{ab}|/\Gamma_{ab}$ for the nine possible Raman transitions from $\z$ to $\p$ as a function of the Raman detuning $\Delta$.  The transition colors and textures follow the same scheme used in Fig. 1: blue and green indicate the observed transitions that conserve the nuclear and total spin, respectively, and red indicates the other transitions, which were not observed.  The vertical black line indicates the approximate detuning used to collect the data shown in Figs. 3 and 4, and the horizontal black lines indicate the calculated ratios for the (upper) $\pN\rightarrow\pN$ and (lower) $\pN\rightarrow\zN$ transitions at that detuning.}
\label{fig.Coherence ratio vs detuning}
\end{center}
\end{figure}

We plot the ratio $|\tilde{\Omega}_{ab}|/\Gamma_{ab}$ in Fig. \ref{fig.Coherence ratio vs detuning}.  As we would expect from our analysis of the toy model, the ratio at $\Delta=0$, when the lasers are tuned between the transitions to $\Ey$ and $\Eone$, is roughly equal to the ratio when the detuning is set far to either side.  We observe that this ratio is substantially reduced when the two Raman lasers come into resonance with transitions to $\Ey$ around +0.33 GHz, to $\Eone$ around -0.33 GHz, and, to a lesser extent, to $\Etwo$ around -2.68 GHz.  The primary limitation on the ratio $|\tilde{\Omega}_{ab}|/\Gamma_{ab}$ stems from this fact, that the off-resonant optical driving that mediates the coherent Raman transition also gives rise to decoherence-inducing optical pumping.

However, we must also contend with pumping on transitions other than those that are necessary to drive the Raman transition, which prevents the ratio from reaching its asymptotic value at very high detunings.  For example, the dip at +1.27 GHz is due to the lower-frequency laser driving the transition from $\z$ to $\Etwo$.  Consequently, we see that the detuning used to collect the data shown in Figs. 3 and 4 is very nearly optimal.  We might gain an increase in coherence by driving the Raman transition at $\Delta=0$, but the gain would be slight and would come at the cost of increased sensitivity to the tuning of our Raman lasers relative to the optical transitions.

We also consider the possibility that the applied magnetic field is not perfectly aligned to the N-V axis.  This misalignment is important because $|E_y^{(0)}\rangle$ and $|E_1^{(0)}\rangle$ are coupled by the Zeeman interaction with the transverse magnetic field in addition to the electronic spin-spin and transverse hyperfine interactions.  For a small misalignment, the transverse Zeeman interaction ($g^{||}_\mathrm{ES}\,\mu_B\,B_\perp/h\approx 190$ MHz for $\theta=5^\circ$, assuming an isotropic $g$-factor), while small compared to the scale of the $\Et$ fine structure, is comparable to or larger than the strengths of the spin-spin ($\lambda_\mathrm{ss}/h=154$ MHz) and hyperfine ($A_\perp^\mathrm{ES}/h=23$ MHz) interactions that mediate the Raman transitions.

\begin{figure}
\begin{center}
\includegraphics[width=\columnwidth]{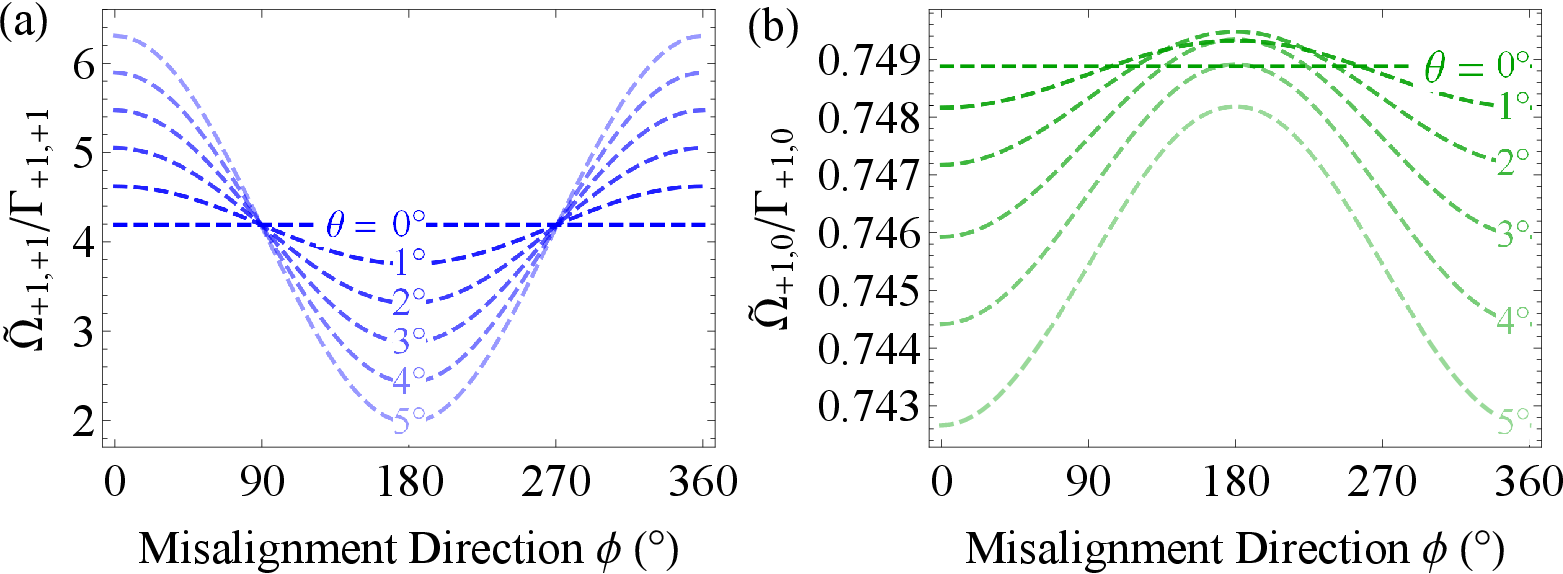}
\caption{Effect of magnetic field misalignment on the coherence of Raman transitions.  We calculate the ratio $|\tilde{\Omega}_{ab}|/\Gamma_{ab}$ for the (a) $\pN\rightarrow\pN$ and (b) $\pN\rightarrow\zN$ transitions, plotted as a function of the angles from the magnetic field to the N-V axis ($\theta$) and to the $x$-axis ($\phi$), which is defined by the $\ExO$ orbital state.}
\label{fig.Coherence ratio vs Bperp}
\end{center}
\end{figure}

Since the Zeeman interaction acts separately on the electronic and nuclear spins, we would expect the presence of a transverse magnetic field to significantly enhance or suppress the strength of the nuclear spin-conserving transitions and to have a smaller, higher-order effect on the electronic-nuclear flip-flop transitions.  In Fig. \ref{fig.Coherence ratio vs Bperp}, we plot the calculated ratio $|\tilde{\Omega}_{ab}|/\Gamma_{ab}$ for the $\pN\rightarrow\pN$ and $\pN\rightarrow\zN$ transitions.  The pumping rates $\Gamma_{ab}$ for both transitions and the Rabi frequency for the $\pN\rightarrow\zN$ transition are mostly unchanged, but a field misalignment of $5^\circ$ can change the Rabi frequency of the $\pN\rightarrow\pN$ transition significantly (by approximately $\pm42$\%).

This result underscores the point that the most important quantity for determining the coherence of the Raman transition is the strength of the interaction that mediates the transition divided by the sum of the decay rates out of the intermediate excited state.  Because the transverse hyperfine interaction mediates the $\pN\rightarrow\pN$ transition but not the $\pN\rightarrow\zN$ transition, a transverse magnetic field can be used to enhance the coherence of the former transition but not of the latter.

\section{Simulation of Coherent Dynamics}
\label{sec.Simulation of Coherent Dynamics}

\subsection{General Approach}
\label{sec.General Approach}

We have calculated the figure of merit $|\tilde{\Omega}_{ab}|/\Gamma_{ab}$ for the various Raman transitions, but we would like to check the results of our model against the observed population dynamics more directly.  To that end, we numerically simulate the dynamics that occur as we drive the two specific Raman transitions studied in the main text.

We consider an 18-level system consisting of the nine ground states as well as the nine excited states corresponding to $\Ey$, $\Eone$, and $\Etwo$.  These simulation basis states are not the electronic-nuclear eigenstates of the full Hamiltonian that are listed in Eq. \ref{eq.Basis States} but rather are the eigenstates of the full Hamiltonian minus the interactions (one component of the spin-spin interaction, the transverse Zeeman interaction, and the transverse hyperfine interaction) that couple $\Ey$ with $\Eone$.  We make the slight simplification that all 18 basis states have well-defined values of $m_I$ and $m_S$ since, due to the strong axial magnetic field, $\Eone$ and $\Etwo$ are respectively polarized into $>99.7\%$ $m_S=+1$ and $m_S=-1$.

We then add the three interactions that couple $\Ey$ and $\Eone$ to each other and the optical transitions that couple them respectively to the $\z$ and $\p$ ground states.  Because every state has well-defined electronic and nuclear spin projections, each ground state is optically coupled to only one excited state and vice versa.  This approach is conceptually analogous to the picture of the toy model shown in \crefformat{figure}{Fig.~#2#1{(a)}#3}\cref{fig.Toy Model}, where the interaction of strength $\lambda$ between the two unmixed excited states is included explicitly.  We numerically simulate the dynamics of the full 18-level system by solving the differential equations from the Heisenberg picture.

Using a common set of input parameters, we perform two simulations: one with $\delta_L$ set to drive the $\pN\rightarrow\pN$ transition and another with $\delta_L$ set to drive the $\pN\rightarrow\zN$ transition.  For each simulation, we assume perfect initial electronic polarization into the $\z$ state, and we use the initial nuclear populations implied by the measurement shown in Fig. 4(a).  Specifically, we extract nuclear populations of 66\% $m_I=+1$, 26\% $m_I=0$, and 8\% $m_I=-1$ from the relative amplitudes of the three nuclear spin-conserving peaks in in the black plot, which shows spectroscopy of the Raman transition performed after nuclear initialization and a 40 $\mu$s wait time.  We then extract the three measured quantities of interest. From the simulation of the $\pN\rightarrow\zN$ transition, we extract the nuclear spin populations of $\pN$ and $\zN$ summed over all three electronic spins.  From the simulation of the $\pN\rightarrow\pN$ transition, we extract the population that has been transferred out of $\z$, summed over $\p$ and $\m$ and all three nuclear states.

Finally, we scale these simulation results to match the nuclear populations measured while driving the $\pN\rightarrow\zN$ transition, as shown in Fig. 4(b), as well as the electronic populations measured while driving the $\pN\rightarrow\pN$ transition, as shown in Fig. 3.  The two measurements were performed on the same day and under the same experimental conditions.  We allow the vertical offset and scaling to vary independently for all three data sets, which accounts for different scalings from population to measured fluorescence.  We set the vertical scaling and offset of the two $\pN\rightarrow\zN$ simulations to match the endpoints of the corresponding data fits (solid purple and orange lines) shown in Fig. 4(b).  We set the vertical offset of the $\pN\rightarrow\pN$ equal to the mean of the grey reference measurement shown in Fig. 3, and we set the scaling to match the endpoint of the data fit at 30 $\mu$s.

\subsection{Simulation Inputs}
\label{sec.Simulation Inputs}

Our simulation uses input values derived from experimental measurements wherever possible.  We extract the energy levels of the ground states from microwave ESR measurements similar to those shown in Sec. \ref{sec.Comparison with Nonresonant Polarization Mechanism} and we extract the energy levels of the excited states from the exact diagonalization of the excited state Hamiltonian described in Sec. \ref{sec.Electronic Hamiltonian}, using the known value of the axial magnetic field and the value of the crystal strain splitting extracted from PLE spectroscopy in Sec. \ref{sec.Electronic Hamiltonian}.  We estimate the Raman detunings $\Delta$ based on the known modulation frequency of 540 MHz that we applied to the EOM to produce the sideband that resonantly drives the $\z\rightarrow\Etwo$ transition during the nuclear polarization stage.  We include the ISC rates described in Sec \ref{sec.Incoherent Optical Pumping Rates} but we neglect the effect of phonon-induced mixing, which would further contribute to the incoherent optical pumping rates.

Because we drive the Raman transition with light that is linearly polarized along the $\EyO$ dipole axis, the ratio of the optical Rabi frequencies $\Omega_{E_y}$ and $\Omega_{E_1}$ of the transitions to $\Ey$ and $\Eone$ is determined by those states' overlap with the $\EyO$ orbital state.  As described in Sec. \ref{sec.Incoherent Optical Pumping Rates}, we find that $|\langle E_y^\prime\Eone|^2\approx0.74$ and $|\langle E_y^\prime\Ey|^2\approx1$, which sets $\Omega_{E_1}/\Omega_{E_y}=0.86$.  We treat the Raman laser power as a free parameter, which we adjust so that the periods of the simulated and observed Rabi oscillations match.

We select the value of $\delta_L$ that resonantly drives a given Raman transition by numerically diagonalizing the full 18-state Hamiltonian, including optical driving, and then calculating the energy difference between the two ground states involved in the transition.  This approach has the advantage of implicitly correcting $\delta_L$ for Stark shifts of the ground states caused by the Raman lasers.

Because all of our basis states have well-defined values of $m_S$ and $m_I$ and because radiative decay should conserve both spin projections, we assume that each excited state radiatively decays solely to the ground state with the same values of $m_S$ and $m_I$.  We assume that the NV center decays with 50\% probability to $\z$ and 25\% probability each to $\p$ and $\m$ after undergoing the ISC into the metastable spin-singlet states, which is consistent with the balanced branching ratio found in Ref. \cite{Robledo2011}.  Because the ISC process from the excited manifold to the spin-singlet states is mediated by electronic spin-orbit coupling and interactions with lattice phonons \cite{Goldman2015a,Thiering2017}, neither of which couples to the nuclear degree of freedom, we assume that the ISC decay pathway conserves $m_I$.

The input parameter with the greatest uncertainty and impact on the simulation results is the strength of the transverse Zeeman interaction, which is related to the degree of magnetic field misalignment.  As discussed in Sec. \ref{sec.Optimizing Parameters}, the application of a magnetic field in certain directions transverse to the N-V axis can significantly enhance or suppress the Rabi frequencies of the nuclear spin-conserving transitions relative to both the Rabi frequencies of the electronic-nuclear flip-flop transitions and the incoherent optical pumping rates.  Instead of parameterizing the transverse Zeeman interaction in terms of the geometric misalignment of the magnetic field, as in Sec. \ref{sec.Optimizing Parameters}, we simply define its strength $\lambda_Z$, which can be either positive or negative to either enhance or suppress the nuclear spin-conserving Rabi frequency.  Therefore, by varying the Raman laser power and $\lambda_Z$, we can independently tune the Rabi frequencies of both transitions.

\subsection{Comparison with Measurement}
\label{sec.Comparison with Measurement}

\begin{figure}
\begin{center}
\includegraphics[width=\columnwidth]{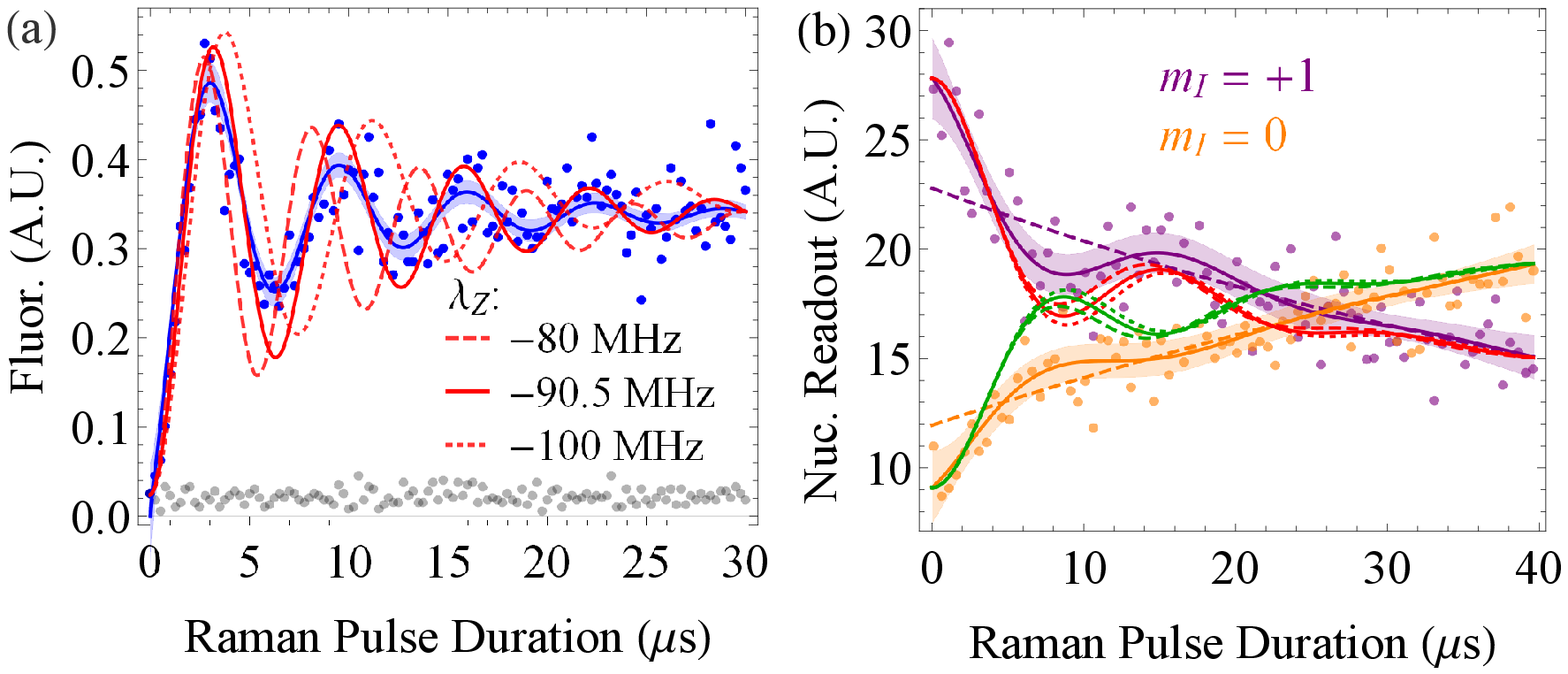}
\caption{Comparison of simulated and measured Raman dynamics corresponding to driving the (a) $\pN\rightarrow\pN$ and (b) $\pN\rightarrow\zN$ transitions.  The three sets of simulation plots [red in (a) and red/green in (b)] correspond to simulations performed using different values of the transverse Zeeman interaction strength $\lambda_Z$.  We show the best fit and 95\% mean prediction interval (solid line and shaded region) of fits to an exponentially damped sinusoid in (a) and an exponentially damped sinusoid plus an exponential ramp in (b).  In (b), we also show the exponential ramp component alone (dashed line).}
\label{fig.Raman Simulation}
\end{center}
\end{figure}

In Fig. \ref{fig.Raman Simulation}, we compare the results of our simulation and fitting procedure to the Raman dynamics measured while driving the $\pN\rightarrow\pN$ and $\pN\rightarrow\zN$ transitions.  As expected, the simulation result for the $\pN\rightarrow\zN$ transition is essentially independent of the value of $\lambda_Z$ used, but varying $\lambda_Z$ does enable us to tune the Rabi frequency of the $\pN\rightarrow\pN$ transition.  We observe close agreement between the simulated and observed Raman dynamics for $\lambda_Z=-90.5$ MHz, which corresponds to a magnetic field misalignment of approximately $5^\circ$ or more.  This close agreement supports our assertion that the coherence with which we can drive the Raman transitions is primarily limited by off-resonant optical pumping by the Raman laser.
	
We can also isolate the oscillatory components of the Raman dynamics observed while driving the $\pN\rightarrow\zN$ transition.  We use the fits shown in Figs. 4(b) and \crefformat{figure}{#2#1{(b)}#3}\cref{fig.Raman Simulation} (solid purple and orange plots), which include both an exponentially damped sinusoidal term and a simple exponential term, and interpret the simple exponential term and the constant offset (dashed purple and orange plots) to represent the population dynamics associated with incoherent pumping.  Subtracting off these terms, we are left with the data shown in Fig. 4(c).  We isolate the oscillatory component of the simulation by fitting the simulation results to the the same fit function and performing the same background subtraction procedure.  Again, the oscillatory behavior of the simulations agrees with that of the measured data.


%